\documentclass[twocolumn,tighten]{aastex631}
\usepackage{color}

\usepackage{multirow}
\usepackage{comment}
\shortauthors{Yamane et al.}

\begin{document}
\title{Associated molecular and atomic clouds with X-ray shell of superbubble 30~Doradus~C in the LMC}

\author[0000-0001-8296-7482]{Y. Yamane}
\affiliation{Department of Physics, Nagoya University, Furo-cho, Chikusa-ku, Nagoya 464-8601, Japan; yamane.y@a.phys.nagoya-u.ac.jp}

\author[0000-0003-2062-5692]{H. Sano}
\affiliation{National Astronomical Observatory of Japan, Mitaka, Tokyo 181-8588, Japan; hidetoshi.sano@nao.ac.jp}

\author[0000-0002-4990-9288]{M. D. Filipovi{\'c}}
\affiliation{Western Sydney University, Locked Bag 1797, Penrith South DC, NSW {2751}, Australia}

\author[0000-0002-2062-1600]{K. Tokuda}
\affiliation{National Astronomical Observatory of Japan, Mitaka, Tokyo 181-8588, Japan; hidetoshi.sano@nao.ac.jp}
\affiliation{Department of Physical Science, Graduate School of Science, Osaka Prefecture University, 1-1 Gakuen-cho, Naka-ku, Sakai 599-8531, Japan}

\author{K. Fujii}
\affiliation{Department of Astronomy, School of Science, The University of Tokyo, 7-3-1 Hongo, Bunkyo-ku, Tokyo 133-0033, Japan}

\author{Y, Babazaki}
\affiliation{Department of Physics, Nagoya University, Furo-cho, Chikusa-ku, Nagoya 464-8601, Japan; yamane.y@a.phys.nagoya-u.ac.jp}

\author{I. Mitsuishi}
\affiliation{Department of Physics, Nagoya University, Furo-cho, Chikusa-ku, Nagoya 464-8601, Japan; yamane.y@a.phys.nagoya-u.ac.jp}

\author{T. Inoue}
\affiliation{Department of Physics, Nagoya University, Furo-cho, Chikusa-ku, Nagoya 464-8601, Japan; yamane.y@a.phys.nagoya-u.ac.jp}

\author[0000-0003-1157-3915]{F. Aharonian}
\affiliation{Max-Planck-Institut für Kernphysik, P.O. Box 103980, D 69029 Heidelberg, Germany}
\affiliation{Dublin Institute for Advanced Studies, School of Cosmic Physics, 31 Fitzwilliam Place, Dublin 2, Ireland}
\affiliation{High Energy Astrophysics Laboratory, RAU, 123 Hovsep Emin St Yerevan 0051, Armenia}

\author{T. Inaba}
\affiliation{Department of Physics, Nagoya University, Furo-cho, Chikusa-ku, Nagoya 464-8601, Japan; yamane.y@a.phys.nagoya-u.ac.jp}

\author[0000-0003-4366-6518]{S. Inutsuka}
\affiliation{Department of Physics, Nagoya University, Furo-cho, Chikusa-ku, Nagoya 464-8601, Japan; yamane.y@a.phys.nagoya-u.ac.jp}

\author[0000-0003-2762-8378]{N. Maxted}
\affiliation{School of Science, University of New South Wales, Australian Defence Force Academy, Canberra, ACT 2600, Australia}

\author{N. Mizuno}
\affiliation{National Astronomical Observatory of Japan, Mitaka, Tokyo 181-8588, Japan; hidetoshi.sano@nao.ac.jp}
\affiliation{Department of Astronomy, School of Science, The University of Tokyo, 7-3-1 Hongo, Bunkyo-ku, Tokyo 133-0033, Japan}

\author[0000-0001-7826-3837]{T. Onishi}
\affiliation{Department of Physical Science, Graduate School of Science, Osaka Prefecture University, 1-1 Gakuen-cho, Naka-ku, Sakai 599-8531, Japan}

\author[0000-0002-9516-1581]{G. Rowell}
\affiliation{School of Physical Sciences, The University of Adelaide, North Terrace, Adelaide, SA 5005, Australia}

\author[0000-0002-2794-4840]{K. Tsuge}
\affiliation{Department of Physics, Nagoya University, Furo-cho, Chikusa-ku, Nagoya 464-8601, Japan; yamane.y@a.phys.nagoya-u.ac.jp}

\author{F. Voisin}
\affiliation{School of Physical Sciences, The University of Adelaide, North Terrace, Adelaide, SA 5005, Australia}

\author[0000-0002-2458-7876]{S. Yoshiike}
\affiliation{Department of Physics, Nagoya University, Furo-cho, Chikusa-ku, Nagoya 464-8601, Japan; yamane.y@a.phys.nagoya-u.ac.jp}

\author{T. Fukuda}
\affiliation{Department of Physics, Nagoya University, Furo-cho, Chikusa-ku, Nagoya 464-8601, Japan; yamane.y@a.phys.nagoya-u.ac.jp}

\author[0000-0001-7813-0380]{A. Kawamura}
\affiliation{National Astronomical Observatory of Japan, Mitaka, Tokyo 181-8588, Japan; hidetoshi.sano@nao.ac.jp}

\author[0000-0003-0890-4920]{A. Bamba}
\affiliation{Department of Physics, School of Science, The University of Tokyo, 7-3-1 Hongo, Bunkyo-ku, Tokyo 133-0033, Japan}
\affiliation{Research Center for the Early Universe, The University of Tokyo, 7-3-1 Hongo, Bunkyo-ku, Tokyo 113-0033, Japan}

\author[0000-0002-1411-5410]{K. Tachihara}
\affiliation{Department of Physics, Nagoya University, Furo-cho, Chikusa-ku, Nagoya 464-8601, Japan; yamane.y@a.phys.nagoya-u.ac.jp}

\author[0000-0002-8966-9856]{Y. Fukui}
\affiliation{Department of Physics, Nagoya University, Furo-cho, Chikusa-ku, Nagoya 464-8601, Japan; yamane.y@a.phys.nagoya-u.ac.jp}
\affiliation{Institute for Advanced Research, Nagoya University, Furo-cho, Chikusa-ku, Nagoya 464-8601, Japan}

\begin{abstract}
30~Doradus~C is a superbubble which emits the brightest {nonthermal} X- and TeV gamma-rays in the Local Group. In order to explore detailed connection between the high energy radiation and the interstellar medium, {we have carried out new CO and H{\sc i} observations using the Atacama Large Millimeter$/$Submillimeter Array (ALMA),  Atacama Submillimeter Telescope Experiment, and the Australia Telescope Compact Array with resolutions of up to 3~pc.} {The} ALMA data of $^{12}$CO($J$~=~1--0) emission {revealed} 23 {molecular} clouds with {the typical diameters} of {$\sim$6--12~pc} and {masses} of $\sim$600--10000 $M_{\odot}$. The comparison with the X-rays of $\it{XMM}$--$\it{Newton}$ at $\sim$3 pc resolution shows that X-rays are enhanced toward these clouds. The CO data were combined with the H{\sc i} to estimate the total interstellar protons. Comparison of the interstellar proton column density and the X-rays revealed that the X-rays are enhanced with the total proton. These are {most likely} due to the shock-cloud interaction modeled by the magnetohydrodynamical simulations \citep{2012ApJ...744...71I}. Further, we {note} a trend that the X-ray photon index varies with distance from the center of the high-mass star cluster, suggesting that the cosmic-ray electrons are accelerated by one or multiple supernovae in the cluster. Based on these results we discuss the role of the interstellar medium in cosmic-ray particle acceleration.
\end{abstract}



\section{Introduction}\label{sec:intro}
30~Doradus~C (hereafter 30~Dor~C) is one of the brightest synchrotron X-ray and TeV gamma-ray superbubbles in the Local Group, which is noticed from the viewpoint of cosmic-ray {acceleration}. \cite{1968MNRAS.139..461L} discovered this superbubble located 200 pc far from the high-mass star-forming region R136 by radio continuum observation. \cite{1984AuJPh..37..321M} detected the shell-like structure in the 843 MHz band with a diameter of $\sim$6$\arcmin$ \citep[$\sim$80 pc at a distance of Large Magellanic Cloud (LMC) of $\sim$50 kpc;][]{2013Natur.495...76P}. The similar shell-like structure was also observed in thermal and non-thermal X-ray emissions \citep[e.g.,][]{1981ApJ...248..925L, 1990ApJ...365..510C, 2001A&A...365L.202D}. Non-thermal X-rays in 30~Dor~C are bright in the northern and western sides of the shell and are confirmed to be synchrotron emission {\citep{2004ApJ...602..257B, 2004ApJ...611..881S, 2009PASJ...61S.175Y, 2015A&A...573A..73K, 2019A&A...621A.138K, 2020ApJ...893..144L}. \cite{2020ApJ...893..144L} performed spectral analysis using {\it XMM}-{\it Newton} and {\it NuSTAR} X-ray data and found the maximum electron energy of $\sim$70--110 TeV, assuming B-field strength of 4 $\mu$G.} In addition, \cite{2015Sci...347..406H} detected gamma-rays from 30~Dor~C whose luminosity is $(0.9 \pm 0.2) \times 10^{35}$ erg s$^{-1}$ at 1--10 TeV. The authors suggested a possibility that the energy of cosmic-rays reach $10^{15.5}$ eV (i.e., ``knee'' energy) in this superbubble. Accordingly, it is certain that VHE cosmic-rays are accelerated in the superbubble 30~Dor~C.


Associated gas with 30~Dor~C has been discovered and this superbubble is known as sites of interaction between the shock and the cloudy medium. \cite{2017ApJ...843...61S} (hereafter Paper I) detected molecular clouds associated with 30~Dor~C using Mopra in the $^{12}$CO$(J=1$--$0)$ transition (beam size id ${\sim}45\arcsec$, corresponding a spatial resolution of $\sim${11} pc at the distance of the LMC). The authors confirmed that most of the molecular clouds are distributed in the western side of the shell. In the western side of the shell, there is OB association LH 90 with ages of 3--4 Myrs and 7--8 Myrs, which is considered to be the parent objects of 30~Dor~C \citep[e.g.,][]{1993A&A...280..426T} and synchrotron X-rays are remarkably bright. Paper I also reported spatial separations of $\sim$10 pc between synchrotron X-ray peaks and CO peaks. They argued that those separations could be interpreted as magnetic field amplification around dense clouds by shock-cloud interaction proposed by \cite{2009ApJ...695..825I, 2012ApJ...744...71I}.

{Subsequently,} \cite{2018ApJ...864...12B} executed a spatially resolved X-ray spectral analysis with a scale of $\sim$10 pc in 30~Dor~C, which led to understanding of the conditions of X-ray emissions (e.g., photon index, intensity, and absorbed column density). The spatial distribution which divided 30~Dor~C into 70 parts shows the spatial anti-correlation between the X-ray photon index and the absorption-corrected intensity, and the tendency that the photon index is small in the western side where the intensity is bright. They compared the distribution of the photon index with that of molecular clouds reported by Paper I qualitatively and suggested a possibility of connection between the variation of the photon index and the shock-cloud interaction {as seen in Galactic SNR and theoretical results \citep[e.g.,][]{2009ApJ...695..825I,2012ApJ...744...71I,2010ApJ...724...59S,2013ApJ...778...59S,2017JHEAp..15....1S,2020ApJ...904L..24S}}. However, {there are no follow up studies} because of a lack of the high-sensitivity and high-resolution CO$/$H{\sc i} data. Therefore, the mechanism responsible for the variation of the X-ray photon index and the intensity was unknown.

In this paper, we present new high-sensitivity and high-resolution data of $^{12}$CO$(J = 1$--0, 3--$2)$ and H{\sc i} emission for the superbubble 30~Dor~C using the Atacama Large Millimeter$/$Submillimeter Array (ALMA), Atacama Submillimeter Telescope Experiment (ASTE), and the Australia Telescope Compact Array (ATCA). By using these data, we intend to reveal the distributions of total interstellar gas (molecular and atomic gas) associated with the superbubble 30~Dor~C with a spatial resolution of $\sim$3--6 pc. In addition, we investigate the mechanism of cosmic-ray acceleration by the interaction of the cloudy medium with the SN shock through a spatial comparison of the interstellar gas with X-ray photon index, intensity, and absorbing column density of. Section \ref{sec:obs} introduces observations and reductions of the data sets we used in this study. Section \ref{sec:res} comprises five subsections: Section \ref{subsec:Overview} describes the spatial distributions of CO, H{\sc i}, and X-ray; Section \ref{physCO} explains the identification of the CO clouds; Section \ref{Vels} gives velocity structures of CO and H{\sc i}; Sections \ref{npdis} and \ref{Xpro} present the results of spatial comparison of the total proton column density with the spectrum conditions in X-rays. In Sections \ref{sec:dis} and \ref{sec:sum} we discuss the results and summarize the conclusions, respectively.

\begin{deluxetable*}{lccccc}[t]
\tabletypesize{\scriptsize}
{
\tablecaption{Observed parameters}
\label{tab:obs}
\tablewidth{0pt}
\tablehead{\\
\multicolumn{1}{c}{Telescopes}&\multicolumn{2}{c}{ALMA}&ASTE&\multicolumn{2}{c}{ATCA}\\
\cline{2-3}\cline{5-6}
&Cycle 3&Cycle 5&&EW352&6B
}
\startdata
Target line&\multicolumn{2}{c}{$^{12}$CO($J$=1--0)}&$^{12}$CO($J$=3--2)&\multicolumn{2}{c}{\ion{H}{1} 21 cm}\\	
Beam size	($''$)&\multicolumn{2}{c}{$\sim$11$ \times15$}&24$\dag$&\multicolumn{2}{c}{16$\dag$}\\
Observing dates&2016$/$Mar.$/$18, 20&2017$/$Dec.$/$22, 29, 30&2015$/$Nov.$/$02, 05, 06&2016$/$Jan.$/$26&2016$/$Jan.$/$28\\
rms noise (K)&$\sim$0.07&$\sim$0.05&$\sim$0.12&\multicolumn{2}{c}{$\sim$8}\\
Base line (m)&7.3--33.7&8.4--35.4&$-$&31--4439&214--5969\\
Velocity resolution (km s$^{-1}$)&\multicolumn{2}{c}{0.4}&0.5&\multicolumn{2}{c}{1.0}\\
Band width&\multicolumn{2}{c}{125 MHz}&128 MHz&\multicolumn{2}{c}{1 GHz}\\
Number of channels (channel)&\multicolumn{2}{c}{2048}&1024&\multicolumn{2}{c}{2048}\\	
Total power&\multicolumn{2}{c}{yes}&$-$&\multicolumn{2}{c}{yes}
\enddata
}
\tablecomments{{$\dag$ Convolved beam size.}}
\vspace*{-0.7cm}
\end{deluxetable*}

\section{Observations and data reduction}\label{sec:obs}
\subsection{{CO and H{\sc i} data}} \label{HI_CO}
{We newly observed CO and \ion{H}{1} data using ALMA, ASTE, and ATCA. Observed parameters are summarized in Table \ref{tab:obs}. We describe the detailed information for each observation as below.}

\subsubsection{ALMA CO $^{12}$CO($J=1$--0)}\label{sub:ALMA}
We carried out ALMA Band 3 (86--116 GHz) observations of $^{12}$CO($J=1$--0) emission line toward the northwest of 30~Dor~C and MC SNR J0536--6913 during the Cycle 3 semester (Proposal $\#$2015.1.01232.S, PI: H. Sano). We used the mosaic$-$mapping mode of a 200$\arcsec$ $\times$ 150$\arcsec$ region centered at ($\alpha_{\rm J2000}, \delta_{\rm J2000}$) $=$ ($5^{\mathrm h}$ $35^{\mathrm m}$ 51$\fs$00, $-69^{\mathrm d}$ $12^{\mathrm m}$ 47$\fs$00) for the northwest of 30~Dor~C, and of a  85$\arcsec$ $\times$ 85$\arcsec$ region centered at ($\alpha_{\rm J2000}, \delta_{\rm J2000}$) $=$ ($5^{\mathrm h}$ $36^{\mathrm m}$ 16$\fs$00, $-69^{\mathrm d}$ $13^{\mathrm m}$ 27$\fs$00) for the MC SNR J0536--6913. The observations used eight 7 m antennas as interferometer and three 12 m antennas as total power array. The observations were executed on March 18 and 20, 2016, through five spectral windows. The target molecular line was $^{12}$CO($J = 1$--0) with a bandwidth of 125.0 MHz (61.035 kHz $\times$ 2048 channels). The baselines ranged from 7.3 m to 33.7 m, corresponding to {\it{u--v}} distances from 2.8 to 12.9 k$\lambda$. The calibration of the complex gains was carried out by observations of J0538--4405, phase calibration on J0601--7036, and flux calibration on J0519--4546.

We also carried out ALMA Band 3 observations of $^{12}$CO($J =1$--0) emission line toward the southwest of 30~Dor~C during the Cycle 5 semester (Proposal $\#$2017.1.01363.S, PI: Y. Yamane). The mosaic-mapping mode was also used to observe 210$\arcsec$ $\times$ 130$\arcsec$ region centered at ($\alpha_{\rm J2000}, \delta_{\rm J2000}$) $=$ ($5^{\mathrm h}$ $35^{\mathrm m}$ 50$\fs$808, $-69^{\mathrm d}$ $12^{\mathrm m}$ 21$\fs$96). The observations used eleven 7 m antennas as interferometer and three 12 m antennas as total power array. The observations were carried out on December 22, 29 and 30, 2017 with {the} same spectral setting of Cycle 3 observations. The baselines ranged from 8.4 to 35.4 m, corresponding to {\it{u--v}} distances from 3.2 to 13.6 k$\lambda$. {The maximum recoverable scale is $\sim$23.7 arcsec for the 7 m array data.} The calibration of the complex gains was carried out using observations of J0522--3627, phase calibration on J0529--7245, and flux calibration on J0516--6207.

For imaging process from the visibility data, we used the Common Astronomy Software Applications package \citep[CASA, version 5.1.0;][]{2007ASPC..376..127M}. We combined 7 m array data taken in both the Cycle 3 and 5 semesters. We applied the natural weighting to recover the weak extended emission. We also applied the multi-scale CLEAN algorithm implemented in CASA \citep[][]{2008ISTSP...2..793C}. The multi-scale CLEAN approach is useful to better restore the extended emission and reduce negative structures. Finally, we combined the cleaned 7 m array data with the total power data using feather task implemented in CASA. The synthesized beam size was $\sim$11$\arcsec$ $\times$ 15$\arcsec$ with a position angle of 66.1 degrees, corresponding a spatial resolution of $\sim$2.7 $\times$ 3.6 pc at the LMC distance $\sim$50 kpc. {The rms noise fluctuations of the data taken in the Cycle 3, and Cycle 5 were $\sim$0.07 K, and $\sim$0.05 K, respectively, with a velocity resolution of 0.4 km s$^{-1}$.}

\subsubsection{ASTE CO $^{12}$CO($J =3$--2)}\label{sub:ASTE}
Observations of $^{12}$CO($J =3$--2) line emission were conducted by the ASTE \citep[][]{2004SPIE.5489..763E} during November 2015 {(Proposal$\#$AC152006, PI: H. Sano)}. We used the on-the-fly mode with Nyquist sampling, and the mapping area was 7.5$\arcmin$ $\times$ 7.5$\arcmin$ centered on the position of ($\alpha_{\rm J2000}, \delta_{\rm J2000}$) $=$ ($5^{\mathrm h}$ $36^{\mathrm m}$ $00^{\mathrm s}$, $-69^{\mathrm d}$ $11^{\mathrm m}$ $42^{\mathrm s}$). The frontend was a 2SB SIS mixer receiver called ``DASH 345.'' The typical system temperature was 250 K in the single side band. The backend was the digital spectrometer, ``MAC'' \citep{2000PASJ...52..785S}, which had a velocity resolution of $\sim$0.11 km s$^{-1}$ per channel and coverage of $\sim$111 km s$^{-1}$ at 350 GHz. The pointing accuracy was checked every 2 hours and the measurements were kept within an error of 2$\arcsec$. The intensity calibration was applied by observing N159W [($\alpha_{\rm J2000}, \delta_{\rm J2000}$) $=$ ($5^{\mathrm h}$ $39^{\mathrm m}$ $37^{\mathrm s}$, $-69^{\mathrm d}$ $45^{\mathrm m}$ $32^{\mathrm s}$)] \citep[][]{2011AJ....141...73M}. The beam size was 22$\arcsec$ and we adopted a grid spacing of 7.5$\arcsec$ in these observations. In this study, data were smoothed with a Gaussian kernel to an effective beam size of 24$\arcsec$. The final noise fluctuations were $\sim$0.12 K at the velocity resolution of 0.5 km s$^{-1}$.


\subsubsection{ATCA H{\sc i}}\label{sub:ATCA}
We used the H{\sc i} 21 cm data which were obtained with ATCA and the Parkes radio telescope. The observations by ATCA were carried out by using two configurations on January 26, 2016 (EW352), and on February 28, 2016 (6B) ({Proposal}$\#$C3070, PI: K{.} Fujii). The baseline ranges of these configurations are from 31 to 4439 m for EW352 and from 214 to 5969 m for 6B. The calibration was carried out by observations of PKS 1934--638 as flux and bandpass calibrator, and PKS 0530--727 as phase calibrator. For our ATCA data reduction, we used Miriad software \citep{1995ASPC...77..433S}. The best H{\sc i} images were obtained using neutral weighting scheme of robust $= 0$ and self-calibration. {To recover the diffuse emission, we combined the H{\sc i} data with the previous H{\sc i} survey data of the whole LMC presented by \cite{2003ApJS..148..473K}, which were obtained using ATCA and the Parkes single-dish telescope. We used IMMERGE task which is feathering together (a linear merging process) the total-power corrected image by \cite{2003ApJS..148..473K} and the present interferometer-only image.} The combined H{\sc i} image has an angular resolution of 12$\farcs$24 $\times$ 10$\farcs$90 with a position angle of 45.2 degrees. In this study, data were smoothed with a Gaussian kernel to an effective beam size of 16$\arcsec$. The final noise fluctuation was $\sim$8 K at the velocity resolution of 1.0 km s$^{-1}$.

\begin{deluxetable*}{ccccccccc}[t]
\tabletypesize{\scriptsize}
\tablecaption{{\it{XMM}}-{\it{Newton}} observations used in the analysis}
\label{tab0}
\tablewidth{0pt}
\tablehead{\\
\multicolumn{1}{c}{Obs. ID}  & R.A. (deg.) & Dec. (deg.) & Obs. Date & \multicolumn{3}{c}{Exposure (ks)} &  Orbit & PI name \\
 & & & & MOS1 & MOS2 & PN &
}
\startdata
0104660301  & 83.866666 & $-$69.269722 & 2000--11--25 & 20.9 & 20.2 & -& 177 & Watson\\
0144530101  & 83.866666 & $-$69.269750 & 2003--05--10 & 47.4 & 47.4 & - & 626 & McCray\\
0406840301  & 83.866666 & $-$69.269750 & 2007--01--17 & 75.3 & 77.1 & 61.1 & 1302 & Haberl\\
0506220101 & 83.867916 & $-$69.270278 & 2008--01--11 & 82.1 & 84.7 & 70.8 & 1482 & Haberl\\
0556350101 & 83.867916 & $-$69.270278 & 2009--01--30 & 80.4 & 83.3 & 66.5 & 1675 & Haberl\\
0601200101 & 83.867916 & $-$69.270278 & 2009--12--12 & 87.6 & 87.3 & 82.5 & 1833 & Haberl\\
0650420101 & 83.867916 & $-$69.270278 & 2010--12--12 & 58.2 & 61.7 & 52.8 & 2016 & Haberl\\
0671080101 & 83.867916 & $-$69.270278 & 2011--12--02 & 68.8 & 70.0 & 64.1 & 2194 & Haberl\\
0690510101 & 83.867916 & $-$69.270278 & 2012--12--11 & 14.7 & 67.2 & 59.5 & 2382 & Haberl\\
0743790101 & 83.867916 & $-$69.270278 & 2014--11--29 & 64.3 & 66.5 & 56.3 & 2742 & Haberl\\
0763620101 & 83.867916 & $-$69.270278 & 2015--11--15 & 62.3 & 62.8 & 57.9 & 2918 & Haberl
\enddata
\tablecomments{All exposure times indicate so called ``Good time interval (GTI)''. The target name for all observations was SN~1987A.}
\end{deluxetable*}

\subsection{XMM--Newton X-rays}\label{XMM}
30~Dor~C was observed by the {\it{XMM}-\it{Newton}} EPIC detectors with a total of 15 pointings as of February 2017. We omitted observations that had an effective exposure time (good time interval; GTI) less than 10 ks, and were not centered on SN 1987A in order to improve the imaging quality. The final datasets used in this paper are listed in Table \ref{tab0}. 
We analyzed both the EPIC-MOS and EPIC-pn by using the HEAsoft version 6.18 and the {\it{XMM}-\it{Newton}} Science Analysis System \citep[SAS][]{2004ASPC..314..759G} version 16.0. For the imaging analysis, we used the {\it{XMM}-\it{Newton}} extended source analysis software \citep[ESAS{:}][]{2008A&A...478..575K}. We applied the standard calibration and filtering routines using the SAS tasks emchain/epchain and ESAS tasks $mos$-$filter/pn$-$filter$. The $pn$-$spectra/mos$-$spectra$ and $pn$-$back/mos$-$back$ tasks were used to create the event, exposure, and quiescent particle background (QPB) background images in the 0.3--1.0 keV, 1.0--2.0 keV, 2.0--7.0 keV, and 0.3--7.0 keV in the energy bands for each detector. Finally, we used $merge\_comp\_xmm$ task to create the mosaicked event, exposure, and QPB background images. The adaptive smoothing task adapt merge was also applied so that the signal to noise ratio is at least 300 with the pixel size of $\sim$2$\arcsec$. The combined mosaic images for each energy band are shown in Figure~1(a).

\subsection{Planck{/}IRAS Dust Opacity}\label{Plank}
We also utilized the archival data sets of the dust opacity at 353 GHz $\tau_{353}$ and dust temperature $T_d$, which were provided by using the combined $Planck{/}IRAS$ data with the graybody fitting \citep[for details, see][]{2014A&A...571A..11P}. The data sets are used to compare with the H{\sc i} data. The angular resolution is 5$\farcm$0 ($\sim$75 pc at the LMC) with a grid spacing of 2$\farcm$4. For a comparison between $\tau_{353}$ and H{\sc i}, we set the resolution and grid size of H{\sc i} to that of $\tau_{353}$. {Note that the beam size of the data is roughly equal to the spatial extent of 30~Dor~C.}

\section{Results}\label{sec:res}

\begin{figure*}
\begin{center}
\includegraphics[width=\linewidth,clip]{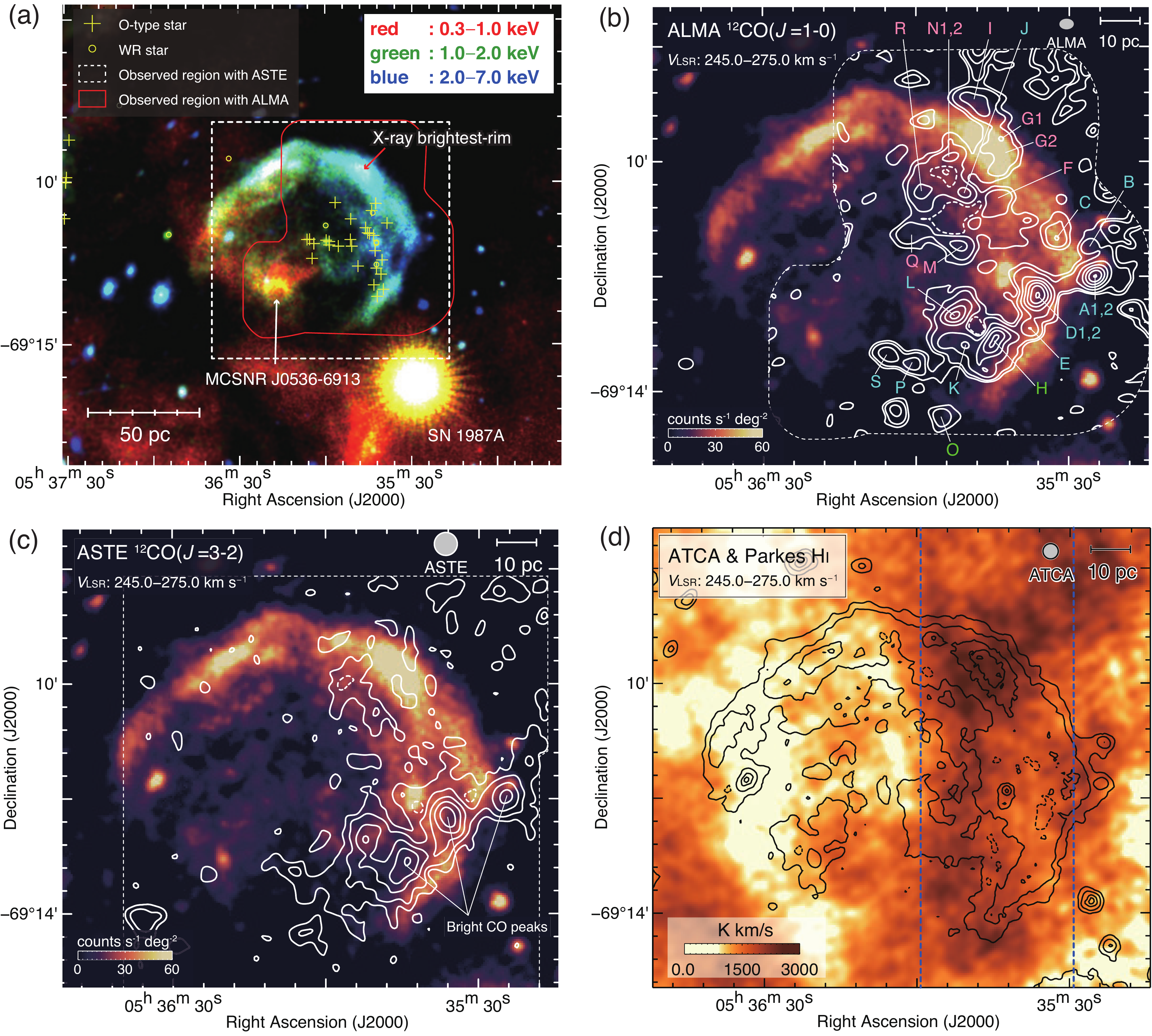}
\caption{(a) Three-color {\it{XMM}-\it{Newton}} EPIC image of 30~Dor~C. Red, green, and blue correspond to the energy bands of  0.3--1.0 keV (soft), 1.0--2.0 keV (medium), and 2.0--7.0 keV (hard), respectively. White dashed line and red solid line show the observed areas of ASTE and ALMA, respectively. Yellow crosses and circles indicate positions of O-type stars and WR stars, respectively. MCSNR J0536$-$6913 and SN 1987A are also shown. (b) Integrated intensity map of ALMA $^{12}$CO$(J$ = 1--0) superposed on the hard X-ray image. The velocity range is from 245 to 275 km s$^{-1}$. The region enclosed by dashed line indicated the observed region with ALMA. Contour levels are 1.03 ($\sim 5$ $\sigma$), 2.07, 5.16, 10.3, 17.6, 26.9, and 38.2 K km s$^{-1}$. The CO clouds A--S are also indicated. The blue, green, and red labels represent molecular clouds whose peak velocity includes velocity ranges of 245--255 km s$^{-1}$, 255--265 km s$^{-1}$, and 265--275 km s$^{-1}$, respectively. (c) Integrated intensity map of ASTE $^{12}$CO$(J$ = 3--2) superposed on the hard X-ray image. The velocity range is the same as in (b). Contour levels are 2.46 ($\sim 12$ $\sigma$), 4.51, 8.26, 13.1, and 18.9 K km s$^{-1}$. The dashed rectangular region indicates the observed region with ASTE. The bright CO peaks are also shown. (d) Integrated intensity map of ATCA \& Parkes H{\sc i} overlaid with the hard X-ray contours. The velocity range is the same as in (b). Contour levels are 5.0, 10, 25, 50, and 85 counts s$^{-1}$ deg$^{-2}$. The vertical dashed lines indicate the integration range of a position-velocity diagram in Figure \ref{fig4}. The resolutions are indicated in the top right corner of each figure.}
\end{center}
\label{fig001}
\end{figure*}%

\subsection{Overview of {CO}, H{\sc i}, and X-ray distributions of 30 Doradus C}\label{subsec:Overview}

We first present the CO, H{\sc i}, and X-ray distributions of the superbubble 30~Dor~C obtained with ASTE, ALMA, ATCA \& Parkes, and {\it{XMM}-\it{Newton}}. Figure~1(a) shows a three-color X-ray image of 30~Dor~C. We found an elliptical shell-like structure of $\sim$80 pc in diameter, which is consistent with the previous X-ray studies \citep[e.g.,][]{2001A&A...365L.202D, 2004ApJ...602..257B, 2015A&A...573A..73K, 2018ApJ...864...12B}. The X-ray shell contains MCSNR J0536--6913 at the position of ($\alpha_{\rm J2000}, \delta_{\rm J2000}$) $\sim$ ($5^{\mathrm h}$ $36^{\mathrm m}$ 17$\fs$0, $-69^{\mathrm d}$ $13^{\mathrm m}$ 28$\fs$0) {\citep{2015A&A...573A..73K, 2017ApJS..230....2B}} . The hard X-rays (blue: 2.0--7.0 keV) are enhanced in the western shell, while the soft X-rays (red: 0.3--1.0 keV) are bright in the eastern shell. The soft X-rays are originated by thermal plasma, whereas the hard X-rays are dominated by the synchrotron radiation \citep[e.g.,][]{2004ApJ...602..257B}. Hereafter, we refer to the soft X-rays as ``thermal X-ray'' and the hard X-rays as ``synchrotron X-rays.'' We also refer to the brightest X-ray peak at ($\alpha_{\rm J2000}, \delta_{\rm J2000}$) $\sim$ ($5^{\mathrm h}$ $35^{\mathrm m}$ 46$\fs$5, $-69^{\mathrm d}$ $9^{\mathrm m}$ 38$\fs$4) as {the} ``X-ray brightest-rim.'' It should be noted that 26 O-type stars and six WR stars are distributed in the western side of 30~Dor~C.

Figure~1(b) shows the distribution of ALMA $^{12}$CO($J =1$--0) superposed with the synchrotron X-ray image obtained with {\it{XMM}-\it{Newton}}. We select the integration velocity range from 245 to 275 km s$^{-1}$, which is roughly consistent with the velocity range to be associated with 30~Dor~C as mentioned by Paper I. The observed area is only the western half of the shell where {$^{12}$CO($J =1$--0)} emission was significantly detected {by Mopra (see Paper I)}. We also show CO clouds A--S that will be detailed in Section \ref{physCO}. We find bright CO peaks toward the southwest of the shell, which likely corresponding to MC1 and MC2 identified by Paper I (CO clouds A1, A2, D1, D2, and H). These CO peaks show a good spatial anti-correlation with the synchrotron X-ray peaks. We also identified the CO clouds toward the X-ray brightest-rim, corresponding to MCs 3--5 identified by Paper I (e.g., CO clouds G1, and G2). Other CO clouds are also located near the synchrotron X-ray shell (e.g., CO clouds B, C, E, F, I--N, and R).

Figure~1(c) shows the integrated intensity contours of the ASTE $^{12}$CO($J =3$--2) superposed on the same X-ray image as shown in Figure 1(b). Molecular clouds appear along the western side of the X-ray shell, whereas there are no dense clouds in the eastern side. We find three bright CO peaks toward the southwest of 30~Dor~C, which are nicely rim-brightened in synchrotron X-rays. We also identified the diffuse CO clouds toward the X-ray brightest-rim. The distribution of $^{12}$CO($J =3$--2) emission is roughly consistent with that of $^{12}$CO($J =1$--0) emission except for the northwestern shell where $^{12}$CO($J =3$--2) emission is weaker than $^{12}$CO($J =1$--0) emission.

Figure~1(d) shows the integrated intensity map of the ATCA \& Parkes H{\sc i} overlaid on the {\it{XMM}-\it{Newton}} X-ray contours. The H{\sc i} emission in the western shell is roughly $\sim$2 times brighter than that in the eastern shell. The X-ray brightest-rim spatially coincides with the H{\sc i} peak named ``NW cloud'' (see also Paper I). No significant absorption features of H{\sc i} spectra are found toward the eastern shell, indicating that the low H{\sc i} brightness temperature means poor atomic gas environment. 

\subsection{Physical properties of the CO clouds}\label{physCO}
\begin{figure}
\begin{center}
\includegraphics[width=\linewidth,clip]{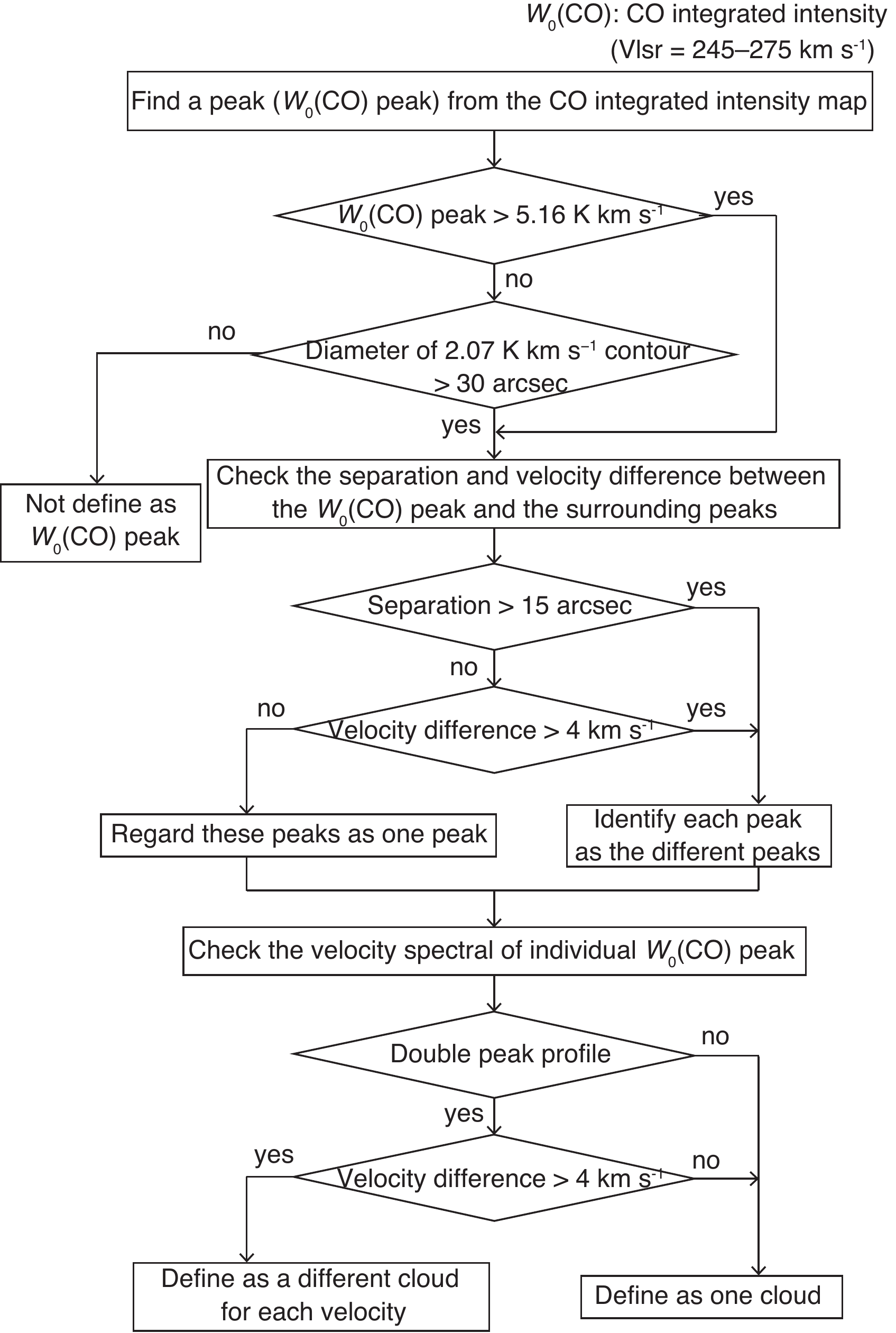}
\caption{{Flowchart of identification of molecular clouds toward 30~Dor~C.}}
\label{fig_new}
\end{center}
\end{figure}

We here present how to derive physical properties of CO clouds. First, we identified CO clouds associated with 30~Dor~C using the ALMA $^{12}$CO($J =1$--0) data. The procedure of defining a CO cloud is as follows {(see also Figure \ref{fig_new})}:

\begin{enumerate}
\item {In the CO integrated intensity map over the whole velocity range 245.0 km s$^{-1}$--275.0 km s$^{-1}$ ($W_0$(CO) see Figure~1(b)), find a peak of $W_0$(CO) greater than 5.16 K km s$^{-1}$ [equal to the 25$\sigma$ level of the ALMA $^{12}$CO($J =1$--0) data] and define it as an individual integrated intensity peak ($W_0$(CO) peak).}
\item {In case that the $W_0$(CO) peak is weaker than 5.16 K km s$^{-1}$, identify it as an individual $W_0$(CO) peak only when the contour of 2.07 K km s$^{-1}$ {[equal to the 10$\sigma$ level]} peak has an equivalent diameter over $\sim$30 arcsec.}
\item {{When the separation between a $W_0$(CO) peak and surrounding peaks is less than 15 arcsec [equal to the beam size of ALMA $^{12}$CO($J =1$--0)] and the velocity difference of these peaks is less than 4 km s$^{-1}$, regard them as one peak.}}
\item {When a $W_0$(CO) {peak} spectrum has a double-peaked profile, each velocity component is considered to belong to different clouds in case that the velocity difference is larger than 4 km s$^{-1}$, $\sim$2.5 times larger than the typical line width of a cloud. We divide the spectra into two components at the velocity of the minimum intensity between the two peaks.}
\item {Peak radiation temperature $T_R^*$, central velocity {$V_\mathrm{peak}$}, and full-width at half maximum line width $\Delta V$ are measured by fitting each spectrum at the $W_0$(CO) peak by a single Gaussian profile.}
\item {Maps of a narrow range integrated intensity ($W_g$(CO)) are made around a each $W_0$(CO) peak whose velocity range is determined from the line width of each spectrum. The surface area $A$ and the position of each cloud are defined as an area enclosed by 5$\sigma$ contours in the $W_g$ (CO) map and the position of the $W_g$ (CO) peak, respectively.}
\end{enumerate}

\begin{deluxetable*}{ccccccrcc}
\tabletypesize{\scriptsize}
\tablecaption{Physical Properties of $^{12}$CO($J$ = 1--0) Clouds}
\label{tab1}
\tablewidth{0pt}
\tablehead{\\
\multicolumn{1}{c}{Name} & $\alpha$(2000) & $\delta$(2000) & $T_R^\ast $ & $V_{\mathrm{peak}}$ &  $\bigtriangleup V$ & \multicolumn{1}{c}{Size}  & \multicolumn{1}{c}{$M_{\mathrm{X}}$} & \multicolumn{1}{c}{$n_{\mathrm{H_2}}$}   \\
 & (h m s) & ($^{\circ}$ $'$ $''$) & (K) & (km $\mathrm{s^{-1}}$) & (km $\mathrm{s^{-1}}$)  & \multicolumn{1}{c}{(pc)} & \multicolumn{1}{c}{(10$^3$ $M_{\sun} $)} & \multicolumn{1}{c}{($10^2$ $\mathrm{cm^{-3}}$)} \\
\multicolumn{1}{c}{(1)} & (2) & (3) & (4) & (5) & (6) & \multicolumn{1}{c}{(7)} & \multicolumn{1}{c}{(8)} & \multicolumn{1}{c}{(9)}
}
\startdata
{A1}	&	5$:$35$:$24.3	&	$-$69$:$11$:$56.3	&	4.3	&	$249.7$	&	3.5	&	9.9	&	4.0	&	3.2	\\
{A2}	&	5$:$35$:$25.1	&	$-$69$:$12$:$02.3	&	7.5	&	$255.2$	&	3.7	&	12.0	&	8.0	&	3.6	\\
{B}	&	5$:$35$:$26.2	&	$-$69$:$11$:$28.3	&	1.8	&	$251.6$	&	1.9	&	5.9	&	0.6	&	2.2	\\
{C}	&	5$:$35$:$32.6	&	$-$69$:$11$:$20.3	&	2.5	&	$250.3$	&	2.1	&	7.8	&	1.1	&	1.9	\\
{D1}	&	5$:$35$:$36.3	&	$-$69$:$12$:$22.3	&	6.3	&	$251.2$	&	4.3	&	13.1	&	10.7	&	3.7	\\
{D2}	&	5$:$35$:$36.0	&	$-$69$:$12$:$18.3	&	5.5	&	$255.9$	&	2.4	&	11.2	&	4.8	&	2.6	\\
{E}	&	5$:$35$:$37.5	&	$-$69$:$12$:$56.3	&	1.9	&	$248.4$	&	3.6	&	7.3	&	1.5	&	3.0	\\
{F}	&	5$:$35$:$43.5	&	$-$69$:$10$:$38.4	&	1.6	&	$268.4$	&	3.1	&	9.8	&	1.9	&	1.6	\\
{G1}	&	5$:$35$:$43.5	&	$-$69$:$09$:$36.4	&	2.5	&	$269.3$	&	1.9	&	12.1	&	2.7	&	1.2	\\
{G2}	&	5$:$35$:$42.0	&	$-$69$:$09$:$52.3	&	2.4	&	$269.9$	&	1.7	&	12.1	&	2.7	&	1.2	\\
{H}	&	5$:$35$:$45.3	&	$-$69$:$13$:$12.4	&	4.3	&	$258.0$	&	5.9	&	12.8	&	10.7	&	4.0	\\
{I}	&	5$:$35$:$46.9	&	$-$69$:$09$:$06.4	&	3.1	&	$269.1$	&	2.2	&	12.9	&	4.6	&	1.6	\\
{J}	&	5$:$35$:$50.6	&	$-$69$:$10$:$32.4	&	4.1	&	$248.2$	&	3.3	&	8.9	&	1.9	&	2.1	\\
{K}	&	5$:$35$:$50.6	&	$-$69$:$13$:$12.4	&	2.4	&	$247.6$	&	4.0	&	11.0	&	4.0	&	2.3	\\
{L}	&	5$:$35$:$52.5	&	$-$69$:$12$:$42.4	&	4.7	&	$247.3$	&	4.7	&	11.0	&	6.7	&	3.9	\\
{M}	&	5$:$35$:$51.4	&	$-$69$:$11$:$28.4	&	1.8	&	$271.3$	&	3.0	&	9.2	&	1.9	&	1.9	\\
{N1}	&	5$:$35$:$52.1	&	$-$69$:$09$:$52.4	&	2.1	&	$266.6$	&	1.9	&	7.7	&	1.0	&	1.7	\\
{N2}	&	5$:$35$:$53.6	&	$-$69$:$09$:$56.4	&	1.6	&	$270.7$	&	1.6	&	9.5	&	1.0	&	0.9	\\
{O}	&	5$:$35$:$54.7	&	$-$69$:$14$:$28.4	&	3.0	&	$257.2$	&	1.5	&	6.5	&	0.8	&	2.3	\\
{P}	&	5$:$36$:$00.4	&	$-$69$:$13$:$32.3	&	2.8	&	$254.8$	&	2.9	&	7.9	&	1.8	&	2.9	\\
{Q}	&	5$:$35$:$58.1	&	$-$69$:$10$:$32.4	&	2.4	&	$273.4$	&	1.9	&	11.1	&	2.2	&	1.2	\\
{R}	&	5$:$36$:$01.1	&	$-$69$:$11$:$14.3	&	1.8	&	$273.1$	&	2.3	&	12.5	&	3.8	&	1.5	\\
{S}	&	5$:$36$:$05.6	&	$-$69$:$13$:$22.3	&	2.4	&	$252.8$	&	3.5	&	9.4	&	3.0	&	2.8	\\
\enddata
\tablecomments{Col. (1): Cloud name. Cols. (2--3): Position of the maximum CO intensity for each velocity component. Cols. (4--6): Physical properties of the $^{12}$CO($J$ = 1--0) emission obtained at each position. Col. (4): Peak radiation temperature, $T_R^{\ast} $. Col. (5): $V_{\mathrm{peak}}$ derived from a single Gaussian fitting. Col. (6): Full-width at half-maximum (FWHM) line width, $\bigtriangleup V$. Col. (7): Cloud size defined as ($A$/$\pi$)$^{0.5}$ $\times 2 $, where $A$ is the total cloud surface area surrounded by the contour of 5 $\sigma$ level of signal over noise ratio (see the text). Col. (8): Mass of the cloud which was derived by using the relationship between the molecular hydrogen column density $N(\mathrm{H_2})$ and the $^{12}$CO($J$ = 1--0) intensity $W(^{12}$CO), $N(\mathrm{H_2})=7.0\times10^{20}[W(^{12}$CO)/(K km s$^{-1}$)](Fukui et al. 2008). Col. (9): The density of hydrogen molecules.} 
\end{deluxetable*}

As a result, we identified individual 23 CO clouds A--S as shown in Figure~1(b). The physical properties of these clouds are summarized in Table \ref{tab1}. $T_R^*$ ranges from 1.6 K to 7.5 K. The typical $\Delta V$ is $\sim$2--6 km s$^{-1}$ and the cloud size is $\sim$5--12 pc. We also calculated the CO-derived mass $M_{\rm {CO}}$ using the following equation:

\begin{figure*}
\begin{center}
\includegraphics[width=\linewidth,clip]{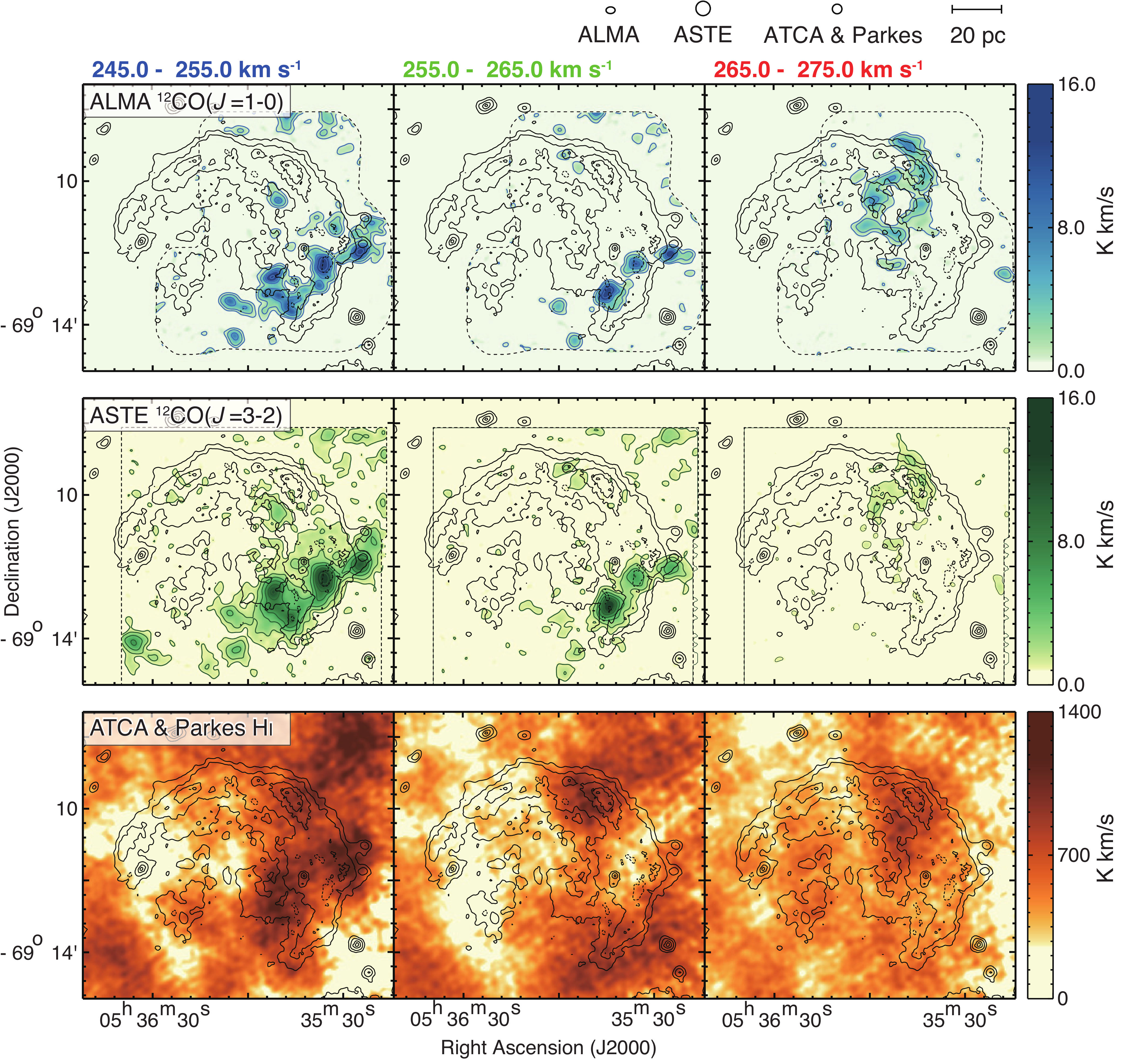}
\caption{Velocity channel maps of ALMA $^{12}$CO$(J$ = 1--0) (top panels), ASTE $^{12}$CO$(J$ = 3--2) (middle panels), and  ATCA \& Parkes H{\sc i} (bottom panels) overlaid with the same X-ray contours as Figure~1(d). Each panel shows the CO and H{\sc i} intensity maps integrated over the velocity range from 245.0 to 275.0 km s$^{-1}$ every 10.0 km s$^{-1}$.  The contour levels are 0.74, 1.92, 5.46, 11.4, and 19.6 K km s $^{-1}$ for $^{12}$CO$(J$ = 1--0); 0.95, 1.90, 3.63, 5.87, 8.53, 11.5, and 14.9 K km s $^{-1}$ for $^{12}$CO$(J$ = 3--2).}
\label{fig2}
\end{center}
\end{figure*}%

\begin{equation}
M_{\mathrm {CO}} = m_{\mathrm H}{\mu} \sum_{i}[D^2 \Omega N_i ({\mathrm {H}_2})],
\label{equ:1}
\end{equation}  
where $m_{\mathrm H}$ is the mass of hydrogen atom, $\mu$ is the mean molecular weight relative to hydrogen atom, $D$ is the distance to the source in unit of cm, equal to 50 kpc, $\Omega$ is the solid angle subtended by a unit grid spacing of a square pixel (0$\farcs$5 $\times$ 0$\farcs$5), and $N_i$(${\mathrm H}_2$) is the hydrogen molecule column density for each pixel in unit of cm$^{-2}$. We adopt $\mu = 2.7$ to take into account the $\sim$36 \% abundance by mass of helium relative to hydrogen molecule. We also used the following relation between the hydrogen molecule column density $N({\mathrm H}_2)$ and the $^{12}$CO($J =1$--0) integrated intensity $W$(CO) from \cite{2008ApJS..178...56F}:

\begin{equation}
N({\mathrm H}_2) = 7.0 \times 10^{20} \cdot W({\mathrm {CO}}) ({\mathrm {cm}}^{-2}),
\label{equ:2}
\end{equation}  
where the units for $W$(CO) are K km s$^{-1}$. Finally, we obtained the typical $M_{\rm {CO}}$ of $\sim$400--9,000 $M_{\sun}$ and the density of $\sim$100--200 cm$^{-3}$ assuming a spherical shape. {Note that the mass, $N({\mathrm H}_2)$, and density for each clouds are thought to have $\sim$30\% relative errors due to uncertainties in {E}quation (\ref{equ:2}) \cite[cf.][]{2013ARA&A..51..207B}.}

\begin{figure*}
\begin{center}
\includegraphics[width=\linewidth,clip]{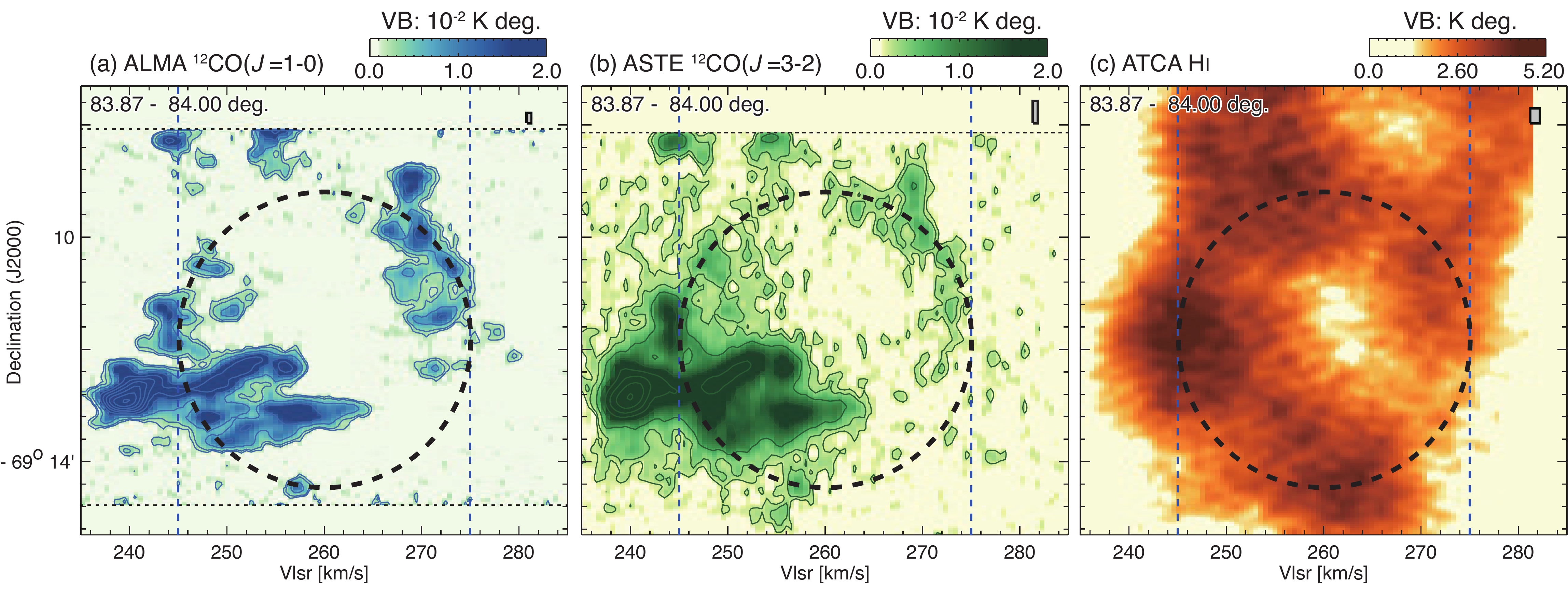}
\caption{Position-velocity diagrams of the (a) ALMA $^{12}$CO$(J$ = 1--0), (b) ASTE $^{12}$CO$(J$ = 3--2), and (c) ATCA $\&$ Parkes H{\sc i}. The integration ranges are from 83$\fdg$87 to 84$\fdg$00 in the Right Ascension as shown in Figure~1(d) by the dashed blue lines. The contour levels are 0.25 {($\sim 5 \sigma$)}, 0.40, 0.85, 1.60, 2.65, 4.00, 5.65, 7.60, 9.85, and 12.4 $\times$ 10$^{-2}$ K deg for $^{12}$CO$(J$ = 1--0); 0.26 {($\sim 8 \sigma$)}, 0.42, 0.88, 1.66, 2.76, 4.16, 5.88, and 7.90 $\times$ 10$^{-2}$ K deg for $^{12}$CO$(J$ = 3--2). The resolutions are indicated in the top right corner of each figure. All black dashed circles are same size and show the shell-like shell toward 30~Dor~C.}
\label{fig4}
\end{center}
\end{figure*}%

\begin{figure*}
\begin{center}
\includegraphics[width=\linewidth,clip]{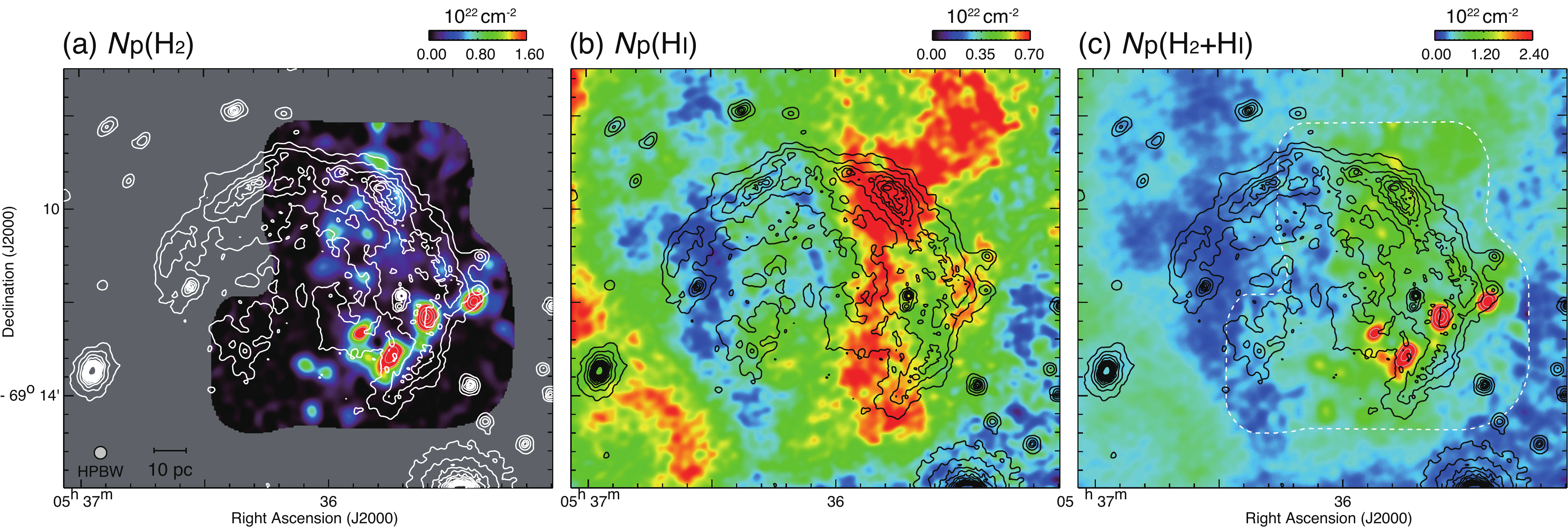}
\caption{Maps of (a) $N_{\rm p}(\mathrm{H}_2)$, (b) $N_{\rm p}$(H{\sc i}), and (c) $N_{\rm p}$($\mathrm{H}_2$+H{\sc i}). The velocity range is from $V_{\rm LSR}$$=$245 to 275 km s$^{-1}$. White and black contours are hard-band {\it{XMM}-\it{Newton}} X-rays delineated 5.00, 10.0, 19.1, 31.0, 45.0, 60.9, and 78.5 counts s$^{-1}$ deg$^{-2}$. White dashed line indicates the observed area of $^{12}$CO$(J$ = 1--0) line emission.}
\vspace*{0.5cm}
\label{fig6}
\end{center}
\end{figure*}%


\subsection{Velocity structures of CO and H{\sc i} Clouds}\label{Vels}
Figure \ref{fig2} shows the velocity channel distributions of $^{12}$CO($J =1$--0), $^{12}$CO($J =3$--2), and H{\sc i} overlaid with the {\it{XMM}-\it{Newton}} X-ray contours. For the ALMA $^{12}$CO($J =1$--0), the CO clouds at $V_{\mathrm {LSR}} = 245$--265 km s$^{-1}$ are located in the southwestern shell, while other molecular clouds at $V_{\mathrm {LSR}} = 265$--275 km s$^{-1}$ are distributed in the northwestern shell. Distributions of the ASTE $^{12}$CO($J =3$--2) are roughly consistent with that of the ALMA $^{12}$CO($J =1$--0), except for the northwestern clouds at $V_{\mathrm {LSR}} = 265$--275 km s$^{-1}$. For the ATCA $\&$ Parkes H{\sc i}, dense atomic clouds are distributed in the western shell. The H{\sc i} clouds tend to be bright in the southwestern shell at $V_{\mathrm {LSR}} = 245$--255 km s$^{-1}$, while the northwestern shell with the X-ray brightest rim is bright in H{\sc i} at $V_{\mathrm {LSR}} = 255$--275 km s$^{-1}$.

Figures \ref{fig4}(a), \ref{fig4}(b), and \ref{fig4}(c) are the position-velocity diagrams of $^{12}$CO($J =1$--0), $^{12}$CO($J =3$--2), and H{\sc i}, whose integration ranges are shown as the dashed lines in Figure~1(d). We confirmed a cavity-like H{\sc i} structure in a velocity range from 245 km s$^{-1}$ to 275 km s$^{-1}$ in Figure \ref{fig4}(c) as previously mentioned by Paper I. In addition to this, we newly find a similar structure of CO, indicated by the dashed black circle in Figures \ref{fig4}(a) and \ref{fig4}(b). The size of CO and H{\sc i} cavity-like structures is roughly consistent with the diameter of 30~Dor~C in Declination.

\begin{figure*}
\begin{center}
\includegraphics[width=\linewidth,clip]{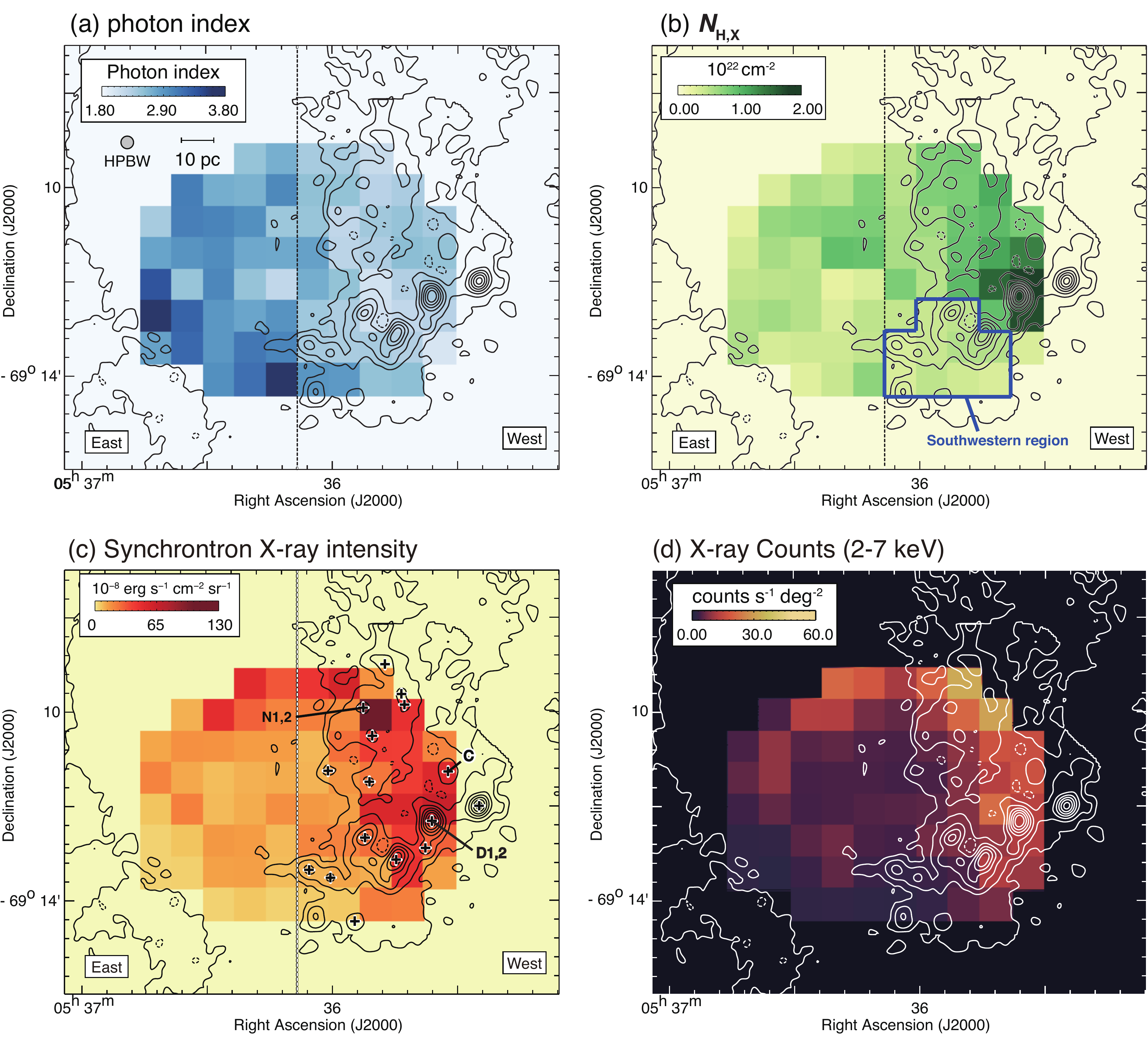}
\caption{Maps of (a) the X-ray photon index, (b) the absorbing column density $N_{\rm{H,X}}$, (c) the absorption-corrected synchrotron X-ray intensity, and (d) the average hard X-ray counts which is the same data as images of {Figures} ~1(b) and ~1(b). {C}ontours indicate $N_{\rm p}$($\mathrm{H}_2$+H{\sc i}) and the contour levels are 0.53, 0.79, 1.27, 2.61, 3.44, 4.35, and 5.35 {$\times 10^{22}$} cm$^{-2}$.}
\label{fig7_2}
\end{center}
\end{figure*}

\subsection{Distribution of proton column density}\label{npdis}
To derive the total proton column density $N_p({\mathrm H}_2+$H{\sc i}) toward 30~Dor~C, we used following equations:

\begin{equation}
N_p({\mathrm H}_2+{\mathrm{H}\textsc{i}}) = 2 \times N_p({\mathrm H}_2) + N_p({\mathrm{H}\textsc{i}}) ({\mathrm {cm}}^{-2}),
\label{equ:3}
\end{equation}  
\begin{equation}
N_p({\mathrm{H}\textsc{i}}) = 1.823 \times 10^{18} \cdot W({\mathrm{H}\textsc{i}}) \cdot X ({\mathrm {cm}}^{-2}),
\label{equ:4}
\end{equation}  
where $N_p({\mathrm H}_2)$ is the proton column density of molecular form, $N_p$(H{\sc i}) is the proton column density of atomic form, $W$(H{\sc i}) is integrated intensity of H{\sc i}, and $X$ is a scaling factor. In general, $X = 1$ was used under the assumption of optically thin H{\sc i} \citep[e.g.,][]{1990ARA&A..28..215D}. However, recent H{\sc i} studies toward the high-latitude clouds indicate that the H{\sc i} is not completely optically thin and $X \sim 1.5$ is better to trace the real amount of atomic hydrogen {\citep[][]{2014ApJ...796...59F, 2015ApJ...798....6F, 2017ApJ...838..132O,2019ApJ...878..131H,2019ApJ...884..130H}.} In the present study, we use $X = 1.7$ (see Appendix).
Figures \ref{fig6}(a), \ref{fig6}(b), and \ref{fig6}(c) are maps of $N_p$(H$_2$), $N_p$(H{\sc i}), and $N_p$(H$_2$+H{\sc i}), respectively. For the map of $N_p$(H$_2$), four prominent structures corresponding to CO clouds A, D, H, and L are seen in the southwestern shell, which have the column density of $\sim$1.6 $\times$ $10^{22}$ cm$^{-2}$ or higher. In the northwestern shell, there is a clumpy structure with column density of $\sim$0.8 $\times 10^{22}$ cm$^{-2}$. For the map of $N_p$(H{\sc i}), the large column density of $\sim$0.6 $\times$ 10$^{22}$ cm$^{-2}$ is seen in the western shell, while the column density of the eastern shell is typically lower than $\sim$0.5 $\times$ 10$^{22}$ cm$^{-2}$. We also note that the highest value of $N_p$(H{\sc i}) $\sim 1.0 \times 10^{22}$ cm$^{-2}$ is placed toward the X-ray brightest-rim. For the map of $N_p$(H$_2$+H{\sc i}), the typical column density of the western shell is $\sim$4 times higher than that of the eastern shell. In the western shell, the total proton column density is dramatically changed by region-to-region with $N_p$(H$_2$+H{\sc i}) $\sim 0.3$--5.4 $\times$ $10^{22}$ {cm$^{-2}$}. By contrast, the eastern half of the shell is expected to show homogeneous gas distribution with $N_p$(H$_2$+H{\sc i}) $ \sim0.4 \times 10^{22}$ cm$^{-2}$ with a small standard deviation of $\sim$0.1$\times 10^{22}$ cm$^{-2}$.

\begin{figure*}
\begin{center}
\includegraphics[width=\linewidth,clip]{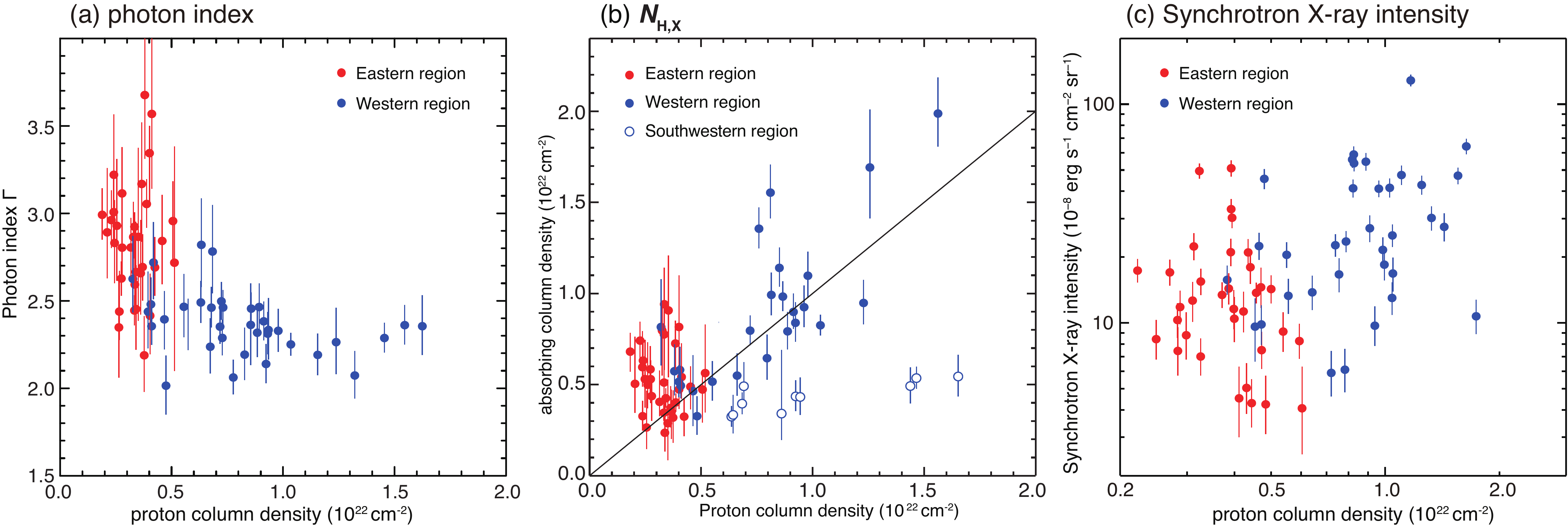}
\caption{Correlation plots for (a) the photon index vs. $N_{\rm p}$($\mathrm{H}_2$+H{\sc i}), (b) $N_{\rm{H,X}}$ vs. $N_{\rm p}$($\mathrm{H}_2$+H{\sc i}), and, (c) the absorption-corrected synchrotron X-ray intensity and $N_{\rm p}$($\mathrm{H}_2 + $H{\sc i}). Eastern and western regions are defined as follow Figure \ref{fig7_2}. (b) Open circles represent the data which were extracted from the southwestern region enclosed by blue line in Figure \ref{fig7_2}(b). Black line shows the proportional relation.}
\label{fig7_3}
\end{center}
\end{figure*}

\subsection{Detailed comparison with the ISM and synchrotron X-ray properties}\label{Xpro}
To test a physical connection between the ISM and X-ray properties, we compare the ISM distribution with the distributions of the photon index, absorbing column density, and synchrotron X-ray intensity presented by \cite{2018ApJ...864...12B}. Their spatially resolved X-ray spectral analysis was carried out with a scale of $\sim$10 pc.
 
Figure \ref{fig7_2} shows a spatial comparison between X-ray properties and ISM. The maps of the photon index, X-ray absorbing column density $N_{\rm{H,X}}$, absorption corrected flux of synchrotron X-rays we used were derived and published by \cite{2018ApJ...864...12B}. These contours indicate $N_p$(H$_2$+H{\sc i}). Figure \ref{fig7_2}(a) represents the photon index {map}. We note that the regions with small photon indices $<$ 2.4 are seen toward the western shell with the dense ISM, while the eastern shell has no dense ISM {and smaller photon indices}. {Figure} \ref{fig7_2}(b) indicates $N_{\rm{H,X}}$. In the northwestern shell, the ISM shows a good spatial correspondence with $N_{\rm{H,X}}$. In the southwestern region indicated as blue box in {Figure} \ref{fig7_2}{(b)}, however, low $N_{\rm{H,X}}$ of $\sim 0.5 \times 10^{22}$ cm$^{-2}$ is found despite the ISM-rich environment around the position of ($\alpha_{\rm J2000}, \delta_{\rm J2000}$) $\sim$ ($5^{\mathrm h}$ $35^{\mathrm m}$ 45$^{\mathrm s}$0, $-69^{\mathrm d}$ $13^{\mathrm m}$ 12$^{\mathrm s}$). Figure \ref{fig7_2}(c) shows the absorption-corrected synchrotron X-ray intensity map. The X-ray intensity is greater than $\sim$1 $\times 10^{-7}$ erg s$^{-1}$ cm$^{-2}$ sr$^{-1}$ toward the eastern shell where the ISM is rich. The X-ray intensity is increased around the ISM peaks (e.g., clouds C, D1, D2, N1, and N2). On the other hand, the northeastern shell is also bright in the synchrotron X-rays where the ISM column density is low. Figure \ref{fig7_2}(d) is the average hard X-ray counts, which is the same data with images of {Figures} ~1(b) and {\ref{fig001}}(c), for each pixel as same as {Figures} \ref{fig7_2}(a)--{\ref{fig7_2}}(c). X-ray counts in the western side of the shell tend to be larger than that in the eastern side. 

Figure \ref{fig7_3} indicates correlation plots between the X-ray properties and $N_p$(H$_2$+H{\sc i}). Figure \ref{fig7_3}(a) shows {a} correlation {plot} between the photon index and $N_p$(H$_2$+H{\sc i}). The photon index tends to be small where the column density is high. A negative-correlation between the photon index and column density is seen. The correlation coefficient is about $-$0.57 for $N_p$(H$_2$+H{\sc i}). Figure \ref{fig7_3}(b) shows a scatter plot between the absorbing column density and $N_p$(H$_2$+H{\sc i}). The absorbing column density is generally proportional to $N_p$(H$_2$+H{\sc i}) except for the three points within the ranges of $N_p$(H$_2$+H{\sc i}) $\sim (1.4$--1.7) $\times 10^{22}$ cm$^{-2}$ and the absorbing column density of $\sim0.5 \times 10^{22}$ cm$^{-2}$. The correlation coefficient is estimated to be $\sim$0.46. Figure \ref{fig7_3}(c) shows a correlation plot between the X-ray intensity and $N_p$(H$_2$+H{\sc i}) {in double logarithm}. We can see a weak positive correlation between the two. We derived the correlation coefficient to be $\sim$0.45.

\section{Discussion}\label{sec:dis}
\subsection{Interstellar gas motion}\label{subsec:gasmo}
Paper I found an H{\sc i} cavity-like structure in the velocity range from 251 to 276 km s$^{-1}$ by using ATCA $\&$ Parkes H{\sc i} data with 1$\arcmin$ resolution. 
{The authors showed that the momentum of these interstellar gas is ten times larger than that of stellar winds. Therefore, Paper I suggested that the lack of the momentum might be derived from pre-existing gas motion.}
In this section, we re-examine the origin of the interstellar gas motion using the new H{\sc i} data with 16$\arcsec$ resolution and ALMA $^{12}$CO$(J=1$--0) data.



In {Figures} \ref{fig4}(a) and {\ref{fig4}}(b), we find two CO components at $V_{\mathrm{LSR}} = 245$--265 km s$^{-1}$ and $V_{\mathrm{LSR}} = 265$--275 km s$^{-1}$, which show the cavity-like structure. We also find a similar structure of H{\sc i} in {Figure} \ref{fig4}(c), which is consistent with the results by Paper I. {Because the Declination extents of CO and H{\sc i} cavity-like structures are roughly consistent with the diameter of X-ray shell, it is natural to think that the cavity-like structures are related to be previous stellar activities of 30~Dor~C: such as strong stellar winds from high-mass stars and clusters and/or past supernova explosions.} In {the present} paper, the velocity range of the structure was determined as $\sim$30 km s$^{-1}$ taking into account the cavity-like structure of not only H{\sc i} but also CO into account, which is different from the velocity range mentioned in Paper I. This difference could arise from the velocity dispersion of $\sim$10 km s$^{-1}$ of the each component. Therefore we assumed that molecular and atomic gas share the same motion in a velocity range of $\sim$30 km s$^{-1}$ in the present study.

First, we consider the momentum of the associated gas motion. Assuming that the CO$/$H{\sc i} cavity-like structure shares the expanding gas motion, the expansion velocity is $\sim$15 km s$^{-1}$. The total mass of the interstellar gas associated with 30~Dor~C is estimated to be $\sim$3.6$\times 10^5 M_{\sun}$, the mass of the molecular gas is $\sim$1.2$\times 10^5 M_{\sun}$, and that of atomic gas is $\sim$2.4$\times 10^5 M_{\sun}$. Consequently, the momentum of the expanding gas motion associated with the superbubble is calculated to be $\sim$5.3$\times 10^6 M_{\sun}$ km s$^{-1}$. On the other hand, the total momentum of the stellar winds is $\sim$2.3$\times 10^5 M_{\sun}$ km s$^{-1}$ from the 26 O-type stars and six WR stars \citep{2009AJ....138.1003B}. 

{Here the momentum of the stellar wind $D_\mathrm{mom}$ is calculated to be $D_\mathrm{mom}=m_\mathrm{loss} \times v \times t$, where $m_\mathrm{loss}$ is the mass loss rate, $v$ is the wind velocity, and $t$ is the wind duration time. The typical wind momentum of high-mass star could be calculated by considering $m_\mathrm{loss}\sim 1 \times 10^{-6} M_{\odot}$ yr$^{-1}$, $v \sim 2000$ km s$^{-1}$, and $t \sim 10^6$ yrs for a O-type star; and $m_\mathrm{loss}\sim 3 \times 10^{-5} M_\mathrm{\odot}$ yr$^{-1}$, $v \sim 2500$ km s$^{-1}$, and $t \sim 3 \times 10^5$ yrs for a WR star \citep[see ][]{1982ApJ...263..723A}.} 

Because stellar winds inject momentum into interstellar gas at $\sim$20$\%$ efficiency \citep{1977ApJ...218..377W}, the effective momentum by stellar winds into associated gas is $\sim$4.6$\times {10^4} M_{\sun}$ km s$^{-1}$, which is only $\sim$1 $\%$ of the momentum of the expanding gas motion.

\begin{figure*}
\begin{center}
\includegraphics[width=\linewidth,clip]{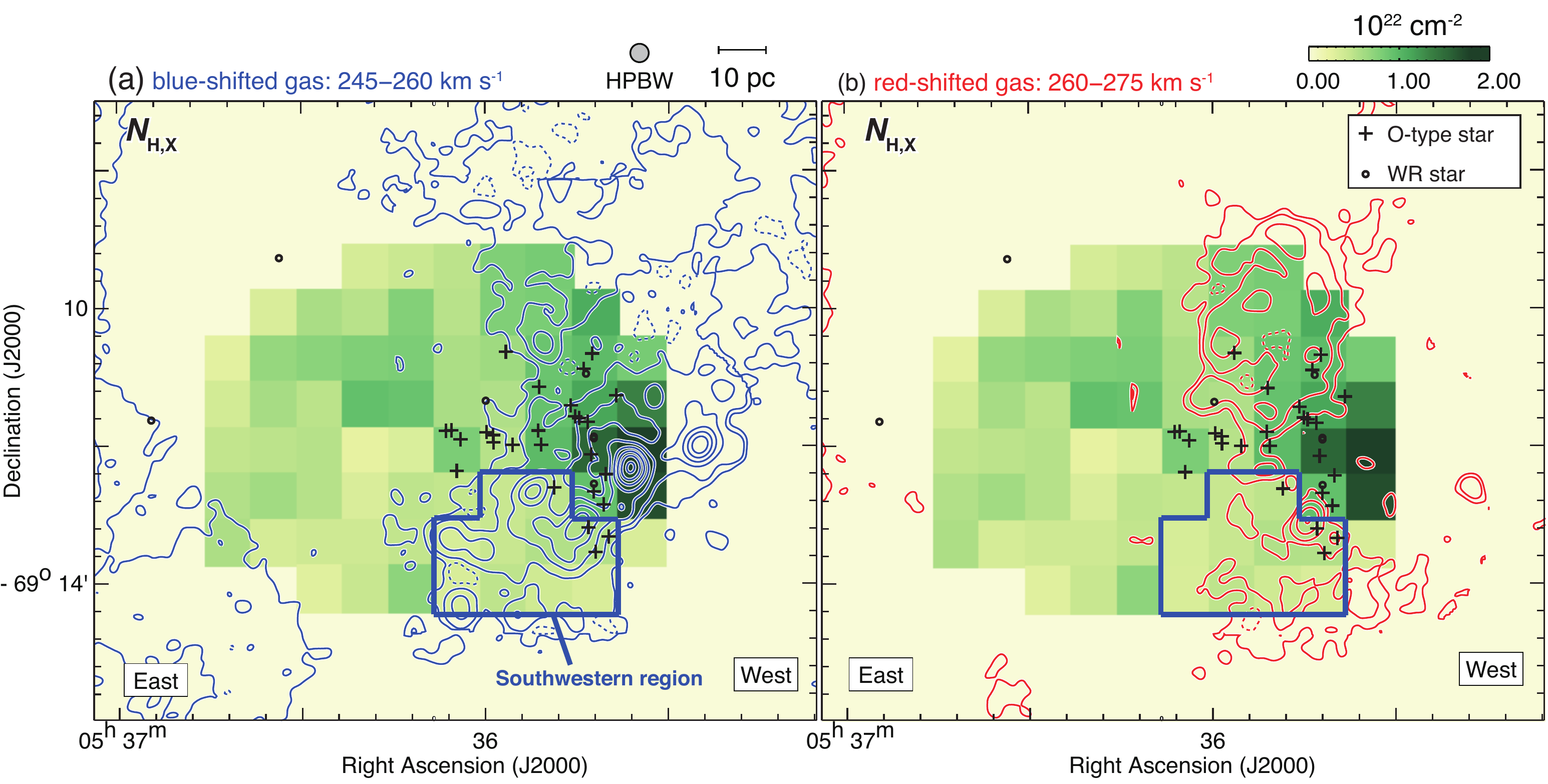}
\caption{Maps of $N_{\rm{H,X}}$ superposed on the contours of total proton column density for the (a) blue-shifted gas, and for the (b) red-shifted gas. The velocity range of the blue-shifted gas is 245--260 km s$^{-1}$, and that of the red-shifted gas is 260--275 km s$^{-1}$. The contour levels are 0.292, 0.414, 0.780, 1.39, 2.24, 3.34, 4.68 $\times10^{22}$ cm$^{-2}$. Black crosses and circles indicate positions of O-type stars and WR stars, respectively.}
\label{fig8_2}
\end{center}
\end{figure*}%

Gas distribution of the line-of-sight also implies the existence of other mechanisms responsible for the formation of the cavity-like structure. Figure \ref{fig7_2}(b) represents the spatial relation between the X-ray absorbing column density and the total proton column density $N_p({\rm H}_2+$H{\sc i}). Figure \ref{fig7_3}(b) shows absorbing column density is roughly proportional to $N_p({\rm H}_2+$H{\sc i}) in many regions, however, in the data points of the southwestern region which plotted as open circles, absorbing column density is small compared $N_p({\rm H}_2+$H{\sc i}). Therefore, we interpret from this relation that in the southwestern region, the interstellar gas has not contributed to the absorption of X-rays, in the other words, the gas is located behind 30~Dor~C. Figure \ref{fig8_2} represents relations between the absorbing column density and $N_p({\rm H}_2+$H{\sc i}) for (a) the blue-shifted gas ($V_{\rm{LSR}} = 245$--260 km s$^{-1}$), and for (b) the red-shifted gas ($V_{\rm{LSR}}  = 260$--275 km s$^{-1}$). Figure \ref{fig8_2}(a) shows that the blue-shifted gas is dominant in the southwestern region, and thus the blue-shifted gas is expected to be located behind 30~Dor~C. On the other hand, {Figure} \ref{fig8_2}(b) shows that the red-shifted gas is located in the region where the absorbing column density is roughly proportional to $N_p({\rm H}_2+$H{\sc i}). Thus, the red-shifted gas is expected to be located in front of 30~Dor~C. In the formation of a cavity-like structure by expanding motion, it is generally thought that the blue-shifted gas is located in front of the red-shifted gas. By contrast, the blue-shifted gas is located behind the red-shifted gas{. This kinematic signature could arise with infall and a collapsing, rather than expanding shell.} Therefore, it is considered that other mechanisms are responsible for the formation of the gas structure whose velocity difference is $\sim$30 km s$^{-1}$.

\begin{figure}
\begin{center}
\includegraphics[width=\linewidth,clip]{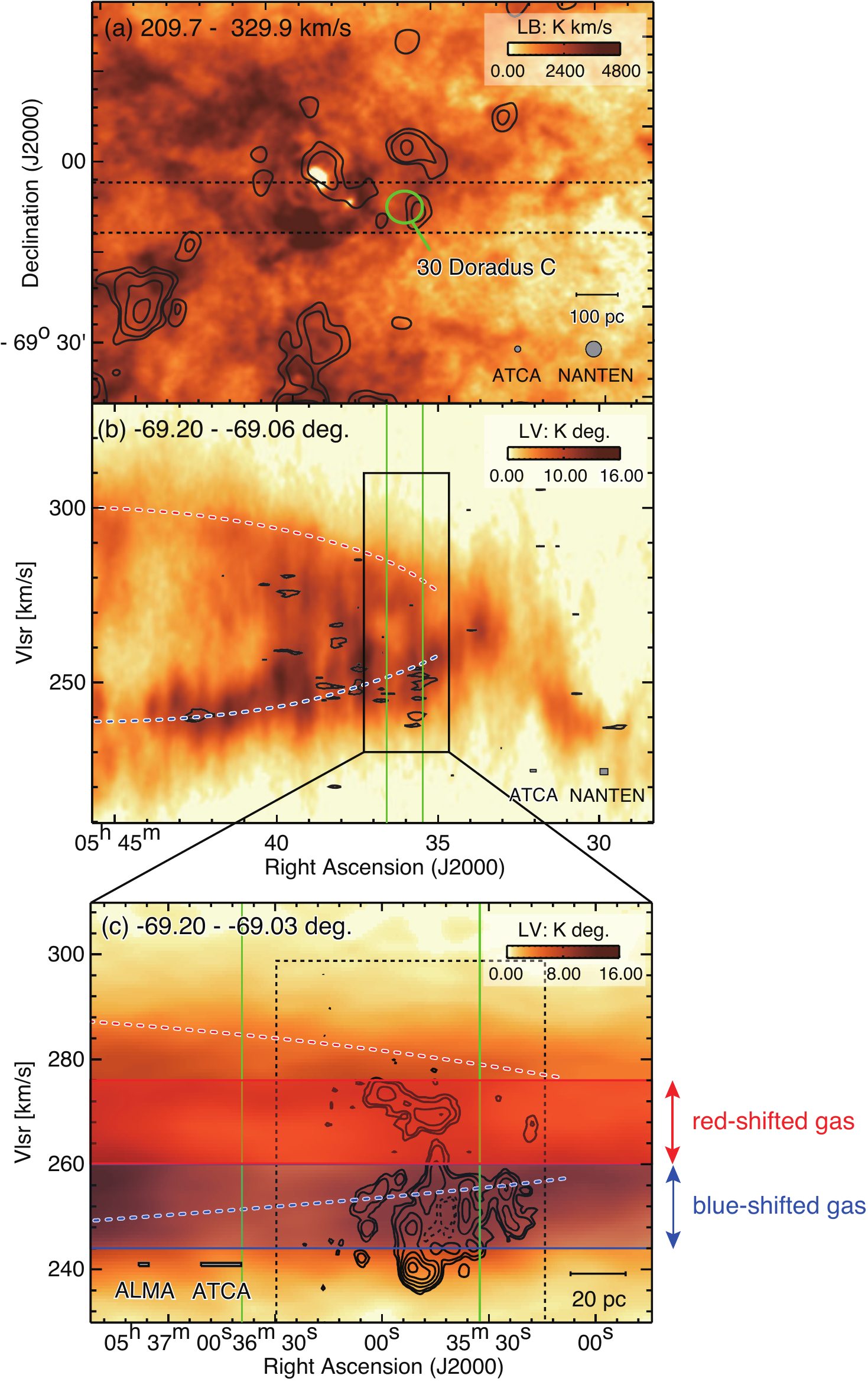}
\caption{(a) Integrated intensity map of H{\sc i} (image) and $^{12}$CO$(J$ = 1--0) (contours) with the integration velocity range of $V_{\rm{LSR}}=209.7$--329.9 km s$^{-1}$. Green circle illustrated the 30~Dor~C and black dashed lines indicate the integration range of (b). The contour levels are 0.859 (5 $\sigma$), 1.72, 4.30, and 8.59 K km s$^{-1}$. (b) Position--velocity diagram of H{\sc i} (image) and $^{12}$CO$(J$ = 1--0) (contours). The position of 30~Dor~C is indicated by the green lines. The integration ranges in Declination are from $-69$\fdg$20$ to $-69$\fdg$03$.  The contour levels are 0.84, and 1.31 $\times$ $10^{-2}$ K km s$^{-1}$. (c) Position--velocity diagram of H{\sc i} and $^{12}$CO$(J$ = 1--0) that enlarges the square region from (b). Red-shifted and blue-shifted gas are corresponding to Figures \ref{fig8_2}(a) and \ref{fig8_2}(b). The contour levels are 0.35, 0.70, 1.75, 3.50, 5.95, and 9.10 $\times$ $10^{-2}$ K km s$^{-1}$.}
\label{fig10}
\end{center}
\end{figure}

We suggest that the gas structure{, large velocity separation of 30 km s$^{-1}$ and unusual kinematic structure,} is driven by galaxy-scale interstellar gas structure as mentioned by Paper I. In the LMC, it has been known that there are two H{\sc i} velocity components whose velocity difference is 50 km s$^{-1}$ \citep[e.g.,][]{1992A&A...263...41L}. \cite{1990PASJ...42..505F} proposed a scenario in which the last tidal interaction between the LMC and Small Magellanic Cloud occurring 0.2 Gyrs ago perturbed the gas in the both galaxies and then formed two velocity components. This galaxy interaction scenario has been supported by numerical simulation by \cite{2007MNRAS.381L..16B}. \cite{2017PASJ...69L...5F} presented the spatial and velocity distribution of these H{\sc i} components in the high-mass star-forming region R136 in the eastern side of {the} LMC. They discovered the bridge feature and complementary distribution between the two H{\sc i} components, which caused the high-mass star formation triggered by a cloud-cloud collision as in the Galactic molecular clouds \citep[e.g., ][]{2018ApJ...859..166F}. Therefore, two H{\sc i} components were interpreted as interacting with each other around R136, and \cite{2017PASJ...69L...5F} concluded that the interaction with two H{\sc i} components triggered the high-mass star formation in and around R136. Likewise, in the star-forming region N44 in the LMC, \cite{2019ApJ...871...44T} presented a scenario of the high-mass star formation triggered by the interaction of the H{\sc i} components. In 30~Dor~C, figure \ref{fig8_2} presents that high-mass stars are located around interstellar gas peaks. In addition, it is considered that the blue-shifted gas approaching from behind the superbubble 30~Dor~C is not involved in the past high-mass star formations because high-mass stars are not distributed in the southwestern region. However, the current resolution is not enough for a further detailed study of star formation in 30~Dor~C. The high-mass star formation will be studied in more detail using ALMA 12-m CO data with higher resolution ($\sim$3$\arcsec$).

Figure \ref{fig10}(a) shows the large-scale integrated intensity map of H{\sc i} around R136 containing 30~Dor~C indicated by green dashed circle, and {Figure} \ref{fig10}(b) shows position-velocity diagram. As noted in {Figure} \ref{fig10}(b), there are two H{\sc i} velocity components which are indicated by red- or blue-dashed lines with the velocity difference $\sim$50 km s$^{-1}$. The velocity difference in the western side is smaller than that in the eastern side. Around 30~Dor~C, two velocity components seem to merge and the velocity difference becomes $\sim$30 km s$^{-1}$. This is in agreement with the velocity difference of the CO$/$H{\sc i} cavity-like structure in 30~Dor~C (see {Figure} \ref{fig4}). In {Figure} \ref{fig10}(c), the velocity ranges of the blue- and red-shifted gas shown in {Figure} \ref{fig8_2} are indicated as red and blue strap, respectively. These two associated gases with 30~Dor~C roughly correspond to two large-scale components indicated by red and blue dashed lines. Like this, the associated gas structure is thought to be related to the large-scale gas structure. {This \ion{H}{1} gas structure suggests the possibility that high-mass star formation is caused triggered by the \ion{H}{1} colliding flow in 30~Dor~C as same as R136 and N44 \citep{2017PASJ...69L...5F, 2019ApJ...871...44T}. In this case, it is likely that coeval high-mass star formation was caused and multiple supernovae exploded in 30~Dor~C.}

{The pre-existing gas motion can also explain the relative positions of red- and blue-shifted clouds which are opposite of what is expected in the expanding motion. \cite{2015ApJ...799..175S} reported the similar trend that the red-shifted CO clouds produced by expanding motion are located on the far side of the Galactic SNR RX~J1713.7$-$3946. The authors discussed that the relative position of the CO clouds is explicable if the pre-existent gas motion cased by overlapping CO supershell GS~347$+$0.0$-$21 is dominant for the CO clouds instead of expansion. In 30 Dor C, the pre-existing gas motion was driven by \ion{H}{1} colliding flow, and the pre-existing gas motion is likely dominant for the expanding gas motion due to the stellar feedback by associated high-mass stars. Note that the spatial scale of the colliding flow is $\sim$1 degree, while that of the expanding gas motion is $\sim$6 arcmin. Therefore, the two kinematic motions are independent but complicated toward 30~Dor~C. I}nvestigating the gas structure around 30~Dor~C using high-resolution {Australian Square Kilometre Array Pathfinder (ASKAP)} H{\sc i} data can better characterize the mechanism responsible for the structure formation.

\subsection{Origin of the synchrotron X-ray}\label{subsec:osync}
30~Dor~C is unique among superbubbles as it has a bright synchrotron X-ray shell. It is known that the maximum energy of accelerated particle in 30~Dor~C is higher than that in the single SNRs \citep[e.g.,][]{2004ApJ...602..257B, 2009PASJ...61S.175Y}. It is important to explore 30~Dor~C for understanding of cosmic-ray acceleration in superbubbles. \cite{2017ApJ...843...61S} focused the distribution of interstellar gas in the northwestern shell and argued that magnetic turbulence is amplified effectively by shock-cloud interaction. On the other hand, \cite{2004PhDT........26B} and \cite{2019A&A...621A.138K} argued that magnetic field strength is lower than 20 $\mu$G. \cite{2019A&A...621A.138K} suggested that, except in the northwestern shell, bright synchrotron X-ray is derived from high energy electrons which are accelerated in the shock front. Based on the size of superbubbles to which the dynamical age corresponds, \cite{2018ApJ...864...12B} suggested that 30~Dor~C is on a phase of high-energy particle acceleration and thus synchrotron X-rays are bright. In this section, we argue that bright synchrotron X-ray emission is derived from the amplification of the magnetic field at the downstream of the shock front by shock-cloud interaction in the western shell with the rich interstellar gas.

First, based on large-scale (10 pc scale) distributions of interstellar gas and synchrotron X-rays, we argue that interstellar gas is related physically to synchrotron X-rays. Both molecular and atomic clouds tend to be distributed at the western side of the shell where synchrotron X-rays are bright (see Figures~1(b)-(d)). In addition, $N_p$(H$_2$+H{\sc i}) in the western side of the shell is $\sim$2--3 times higher than that in the eastern side (see Figure \ref{fig6}). Consequently, we consider that some mechanisms are responsible for an increase of the synchrotron X-ray intensity in such cloud-rich regions. This physical relation seems to have appeared quantitatively as the positive correlation between the proton column density and the absorption-corrected synchrotron X-ray intensity (see Figure \ref{fig7_2}(b), correlation coefficient $\sim$0.40).

Next, based on small-scale (a few pc scale) distributions of CO clouds and synchrotron X-rays, we argue that the relation of the two may be driven by shock-cloud interaction. As shown in Figure~1(b), it is found that all bright synchrotron X-ray spots are associated with CO clouds (Clouds A1, A2, B, C, D1, D2, E, F, G1, G2, I, J, L, N1, and N2) in the western shell, and the separations between each peak are a few pc. These results are considered to be due to the SN shock interacting with the interstellar gas, and which is then amplifying synchrotron X-ray via intensifying a magnetic field around the shocked gas. In fact, these relations between the interstellar gas and X-rays were observed in the Galactic single SNRs, (e.g., RX~J1713.7$-$3946, and RCW~86). The result was interpreted as the turbulence amplification of magnetic field {up to mG} around the dense cloud cores \citep[e.g.,][]{2015ApJ...799..175S, 2017JHEAp..15....1S}. {Interestingly,} the typical separation between the peaks of interstellar gas and X-rays in 30~Dor~C is larger than the average separations in the Galactic SNRs which are $1.2 \pm 0.6$ pc for RX~J1713.7$-$3946, and $1.8 \pm 1.3$ pc for RCW 86. Therefore, it is likely that there is smaller structure in 30~Dor~C as same as that in Galactic SNRs. It is expected that more detailed structure will be revealed by using new ALMA 12 m array data with high-resolution.

{Although the strong magnetic field is expected under the shock-cloud interaction scenario in 30~Dor~C, }\cite{2019A&A...621A.138K} claimed {low magnetic field strength. The} authors made radial profiles of X-rays in the synchrotron shell in the northwest (NW), southwest, and northeast, and made model fitting for filamentary X-rays. They found that the post shock magnetic field strength is less than {20 $\mu$G}. In the analysis they adopted a width greater than 10 pc in the azimuthal direction. The width is 20 times larger than that of the typical filamentary distribution \citep{2005ApJ...621..793B}, making it impossible to resolve the enhanced magnetic field of sub-pc scale \citep[e.g., see RXJ1713,][]{2010ApJ...724...59S, 2013ApJ...778...59S}. Additionally, the scale of the width is two orders larger than the scale of magnetic field amplification we consider \citep[][]{2007Natur.449..576U, 2009ApJ...695..825I}. The authors noted that the fitting was not successful in the regions of molecular gas (sectors 6 and 7 in the NW). The argument that the shock-cloud interaction does not affect the magnetic filed and X-rays in the LMC \citep{2019A&A...621A.138K} is therefore not justified, and measurements of the magnetic field in 30~Dor~C require careful re-examination.

\begin{figure*}
\begin{center}
\includegraphics[width=\linewidth,clip]{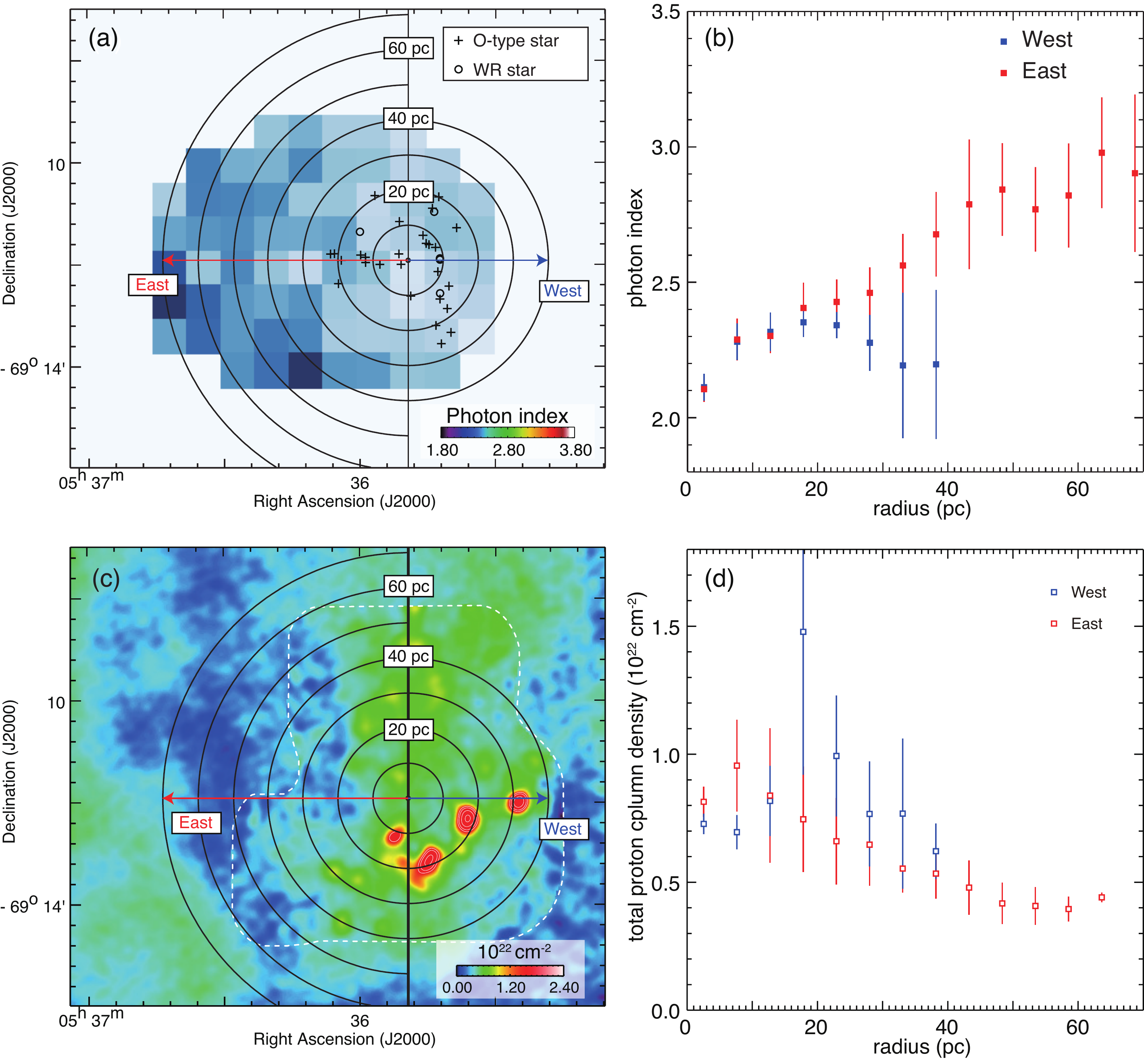}
\caption{(a) Map of the photon index as same as Figures \ref{fig7_2}(a) image. Black crosses and circles indicate O-type stars and WR stars, respectively. (b) Western (blue) and eastern (red) radial profiles of the photon index centered on the center of mass of the high-mass stars. (c) Map of $N_p$(H$_2+$H{\sc i}) as same as Figure \ref{fig6}(c) image. (d) Western (blue) and eastern (red) radial profiles of $N_p$(H$_2+$H{\sc i}) centered on the center of mass of the high-mass stars.}
\label{fig11}
\end{center}
\vspace*{0.5cm}
\end{figure*}%

\subsection{Efficient acceleration of cosmic-ray electrons}\label{subsec:cracc}
{\cite{2018ApJ...864...12B} presented the spatial variation of the photon index between eastern and western parts of the shell in 30~Dor~C.} {The authors found mainly two observational trends: (1) small photon indices are seen in the western part where dense molecular clouds are associated, and (2) absorption-corrected synchrotron X-ray intensity shows a negative correlation with the photon indices. The authors concluded that these trends could be understand as magnetic field amplification around the shocked clouds via shock-cloud interaction \citep[e.g.,][]{2009ApJ...695..825I,2012ApJ...744...71I, 2010ApJ...724...59S,2013ApJ...778...59S}. However, a detailed spatial comparison between interacting clouds and X-ray properties has never been presented.} In this section, we discuss the effective cosmic-ray acceleration and the amplified turbulence in 30~Dor~C based on the spatial distributions of the interstellar gas and the photon index.

We argue that cosmic-ray electrons are accelerated effectively {at least} a few TeV at the region where the photon index is small ($<$ 2.4) presented by \cite{2018ApJ...864...12B}. In the regime of Bohm diffusion, the energy flux of synchrotron X-rays from accelerated electrons has cut-off power-law distributions through the classical diffusive shock acceleration mechanism. It is considered that small (large) photon index of synchrotron X-rays is corresponding to high (low) maximum energy of accelerated electron (roll-off energy), when the energy distribution is fitted by the single power-law model. This is a useful interpretation in the case of a lack of photon statistics, in other words, when the energy distribution cannot be fitted by cut-off power law model. Based on the same interpretation, \cite{2015ApJ...799..175S} presented the spots where high energy electron is accelerated. They divided the Galactic SNR {RX~J1713.7$-$3946} into more than 300 regions and performed the same analysis as \cite{2018ApJ...864...12B}.

The roll-off energy of synchrotron X-ray from accelerated electron in the regime of Bohm diffusion is described as \citep{2007A&A...465..695Z}
\begin{equation}
\varepsilon_0=0.55 \times \left(\frac{v_\mathrm{sh}}{3000 \mathrm{\;km\;s}^{-1}}\right)^2 \eta^{-1}\;\;  \mathrm{(keV)}.
\label{equ:05}
\end{equation}  
Here $v_\mathrm{sh}$ is the shock speed. $\eta$ ($>$1) is gyro factor which represents the degree of turbulence in the upstream. In the Bohm limit, $\eta$ is close to 1 due to a strong magnetic turbulence. As can be seen from {E}quation {(\ref{equ:05})}, the roll-off energy depends on the shock speed and the turbulence in the upstream of the shock. {Hereafter, we examine the effect of both the shock velocity and gyro factor.}

{\subsubsection{Case~1: shock velocity variation}}
We discuss the effects of SN explosions to evaluate the variation of shock speed $v_{sh}$ in {E}quation {(\ref{equ:05})}. First, we {predict} that the center of the last SN explosion would be in {the} western side of the shell because high-mass stars are be located in the western shell. In this case, {the shock velocity of the explosion center (or the western half of the shell) is faster than that of the eastern half of 30~Dor~C,} and thus, {a large roll-off energy (or small photon index) is expected in the western shell} according to {E}quation {(\ref{equ:05})}. To confirm this relation more quantitatively, we plotted east--west radial profiles of photon {indices} centered on the center of mass of the high-mass star cluster, as shown in {Figures \ref{fig11}(a) and \ref{fig11}(b)}. These profiles show that photon {indices tend} to be small near the center of {the high-mass star cluster}. These results suggest that single or multiple SNe occurred in the high-mass star cluster {which accelerated cosmic-ray electrons via diffusive shock acceleration.}\\

{\subsubsection{Case~2: shock-cloud interaction induced variation of gyro factor}}
{We also argue that the shock-cloud interaction may affect the spatial variation of photon indices.} {Figures \ref{fig11}(c) and \ref{fig11}(d)} indicates eastern and western radial profiles of $N_p$(H$_2+$H{\sc i}). We found that  $N_p$(H$_2+$H{\sc i}) is small in the region where photon index is large {(Figures \ref{fig11}}. This is {the} same trend as the results in Figures \ref{fig7_2}(a) and \ref{fig7_3}(a). If this relation is true, it is considered that the roll-off energy is extended by strong magnetic turbulence (that is, $\eta$ in {E}quation {(\ref{equ:05})} is close to 1) around the interstellar gas due to shock-cloud interactions mentioned in Section \ref{subsec:osync}. However, the relation between photon index and interstellar gas is unclear at this time, because spatial resolution of photon index is  $\sim$$10\times10$ pc that is not enough to reveal the connection. We expect that the relation will be understood by a deep $Chandra$ observation which can show the more detail{ed spatial} variation of distribution of the photon index in 30~Dor~C.\\

\section{Summary}\label{sec:sum}
We carried out new high-resolution study of the interstellar medium toward the LMC superbubble 30~Dor~C. We explore the role of the interaction between the shock of an expanding bubble and the interstellar medium and its role in the high energy radiation, in particular non-thermal X-rays. The main conclusions of the study are summarized as follows;

\begin{enumerate}
\item {The new high-resolution of ALMA and ASTE allowed us to reveal detailed distribution{s} of the 23 CO clouds with radius of $\sim$3--6 pc. A comparison of the CO distribution with the non-thermal X-rays of $\it{XMM}$--$Newton$ revealed the enhanced X-rays around the CO clouds. This is explained as due to the shock{-}cloud interaction which was numerical simulated by MHD \citep{2012ApJ...744...71I}. }
\item {The total column density of the interstellar protons in atomic and molecular form are compared with the X-ray photon index. We see a trend that the photon index tends to become small toward higher column density, while scattering is large. This is also consistent with the shock-cloud interaction which produces higher energy cosmic-ray electrons.}
\item {In addition, the X-ray photon index tends to become small toward the center of the high-mass star cluster, suggesting that the high-mass star cluster is responsible for the production of the cosmic-ray electrons via SNe. It is most likely that single or multiple SNe occurred in the cluster {which accelerated cosmic-ray electrons via diffusive shock acceleration.}}
\item {\cite{2017PASJ...69L...5F} presented a scenario that the H{\sc i} colliding flow triggered the formation of R136 and the other high-mass stars surrounding. The H{\sc i} velocity distribution shown in the present work suggests that 30~Dor~C is located toward the merging area of the low velocity H{\sc i}. The high-mass stars in 30~Dor~C are part of these stars, suggesting that their formation is coeval in the last few Myrs. The formation may explain the multiple SNe suggested in Section \ref{subsec:gasmo}.}
\end{enumerate}

\begin{acknowledgements}
We are grateful to the anonymous referee for useful comments which helped the authors to improve the paper. This paper makes use of the following ALMA data: ADS/JAO.ALMA\#2015.1.01232.S and ADS/JAO.ALMA\#2017.1.01363.S. ALMA is a partnership of ESO (representing its member states), NSF (USA) and NINS (Japan), together with NRC (Canada), MOST and ASIAA (Taiwan), and KASI (Republic of Korea), in cooperation with the Republic of Chile. The Joint ALMA Observatory is operated by ESO, AUI/NRAO and NAOJ. This work made use of the {\it{XMM-Newton}} Extended Source Analysis Software. The Australia Telescope Compact Array, and Parkes radio telescope are part of the ATNF which is funded by the Australian Government for operation as a National Facility managed by CSIRO.We acknowledge Kevin Grieve for his valuable support during the H{\sc i} observations and data reduction. The ASTE telescope is operated by National Astronomical Observatory of Japan (NAOJ). This study was financially supported by Grants-in-Aid for Scientific Research (KAKENHI) of the Japanese Society for the Promotion of Science (JSPS, grant Nos. \href{https://kaken.nii.ac.jp/en/grant/KAKENHI-PROJECT-15H05694/}{JP15H05694}, \href{https://kaken.nii.ac.jp/en/grant/KAKENHI-PROJECT-16K17664/}{JP16K17664}, \href{https://kaken.nii.ac.jp/en/grant/KAKENHI-PUBLICLY-19H05075/}{JP19H05075}, and \href{https://kaken.nii.ac.jp/en/grant/KAKENHI-PROJECT-19K14758/}{JP19K14758}. K. Tokuda was supported by NAOJ ALMA Scientific Research Grant Number 2016-03B.
\end{acknowledgements}

\software{CASA \citep [v 5.1.0;][]{2007ASPC..376..127M}, MIRIAD \citep{1995ASPC...77..433S}, SAS \citep{2004ASPC..314..759G}, ESAS \citep{2008A&A...478..575K}}

\facilities{Atacama Large Millimeter$/$Submillimeter Array (ALMA),  Atacama Submillimeter Telescope Experiment (ASTE), Australia Telescope Compact Array (ATCA), Parkes, {\it{XMM-Newton}}}

\begin{figure*}
\begin{center}
\includegraphics[width=\linewidth,clip]{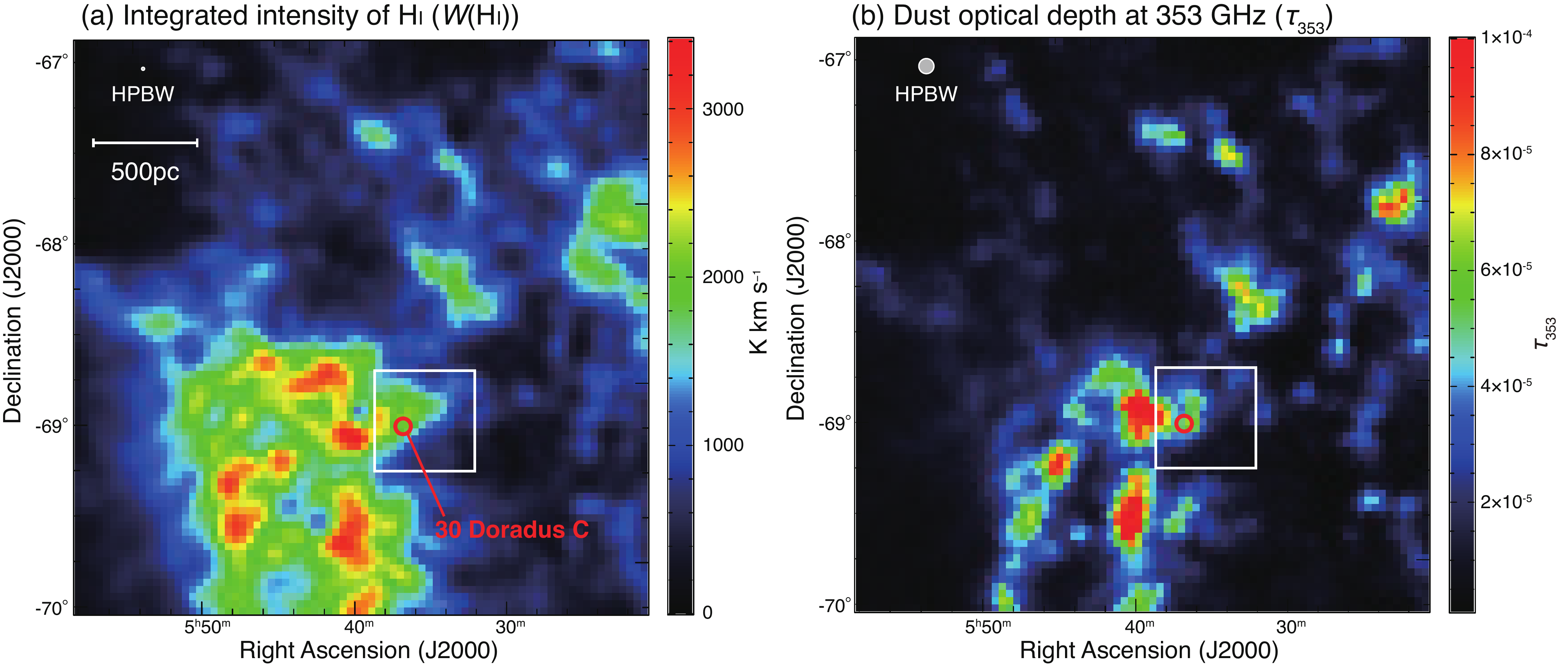}
\caption{Maps of (a) integrated intensity of H{\sc i} ($W$(H{\sc i})) and (b) the dust optical depth at 353 GHz ($\tau$$_{353}$) around 30~Dor~C. The red circles indicate the positions of 30~Dor~C. {The white box is the 30$'\times$30$'$ region centered at ($\alpha_{\rm J2000}$,$\delta_{\rm J2000}$)= ($5^{\mathrm h}$ $34^{\mathrm m}$ 50$^{\mathrm s}$0, $-69^{\mathrm d}$ $13^{\mathrm m}$ 49$^{\mathrm s}$).}}
\label{fig12}
\end{center}
\vspace*{0.5cm}
\end{figure*}%

\begin{figure}
\begin{center}
\includegraphics[width=\linewidth,clip]{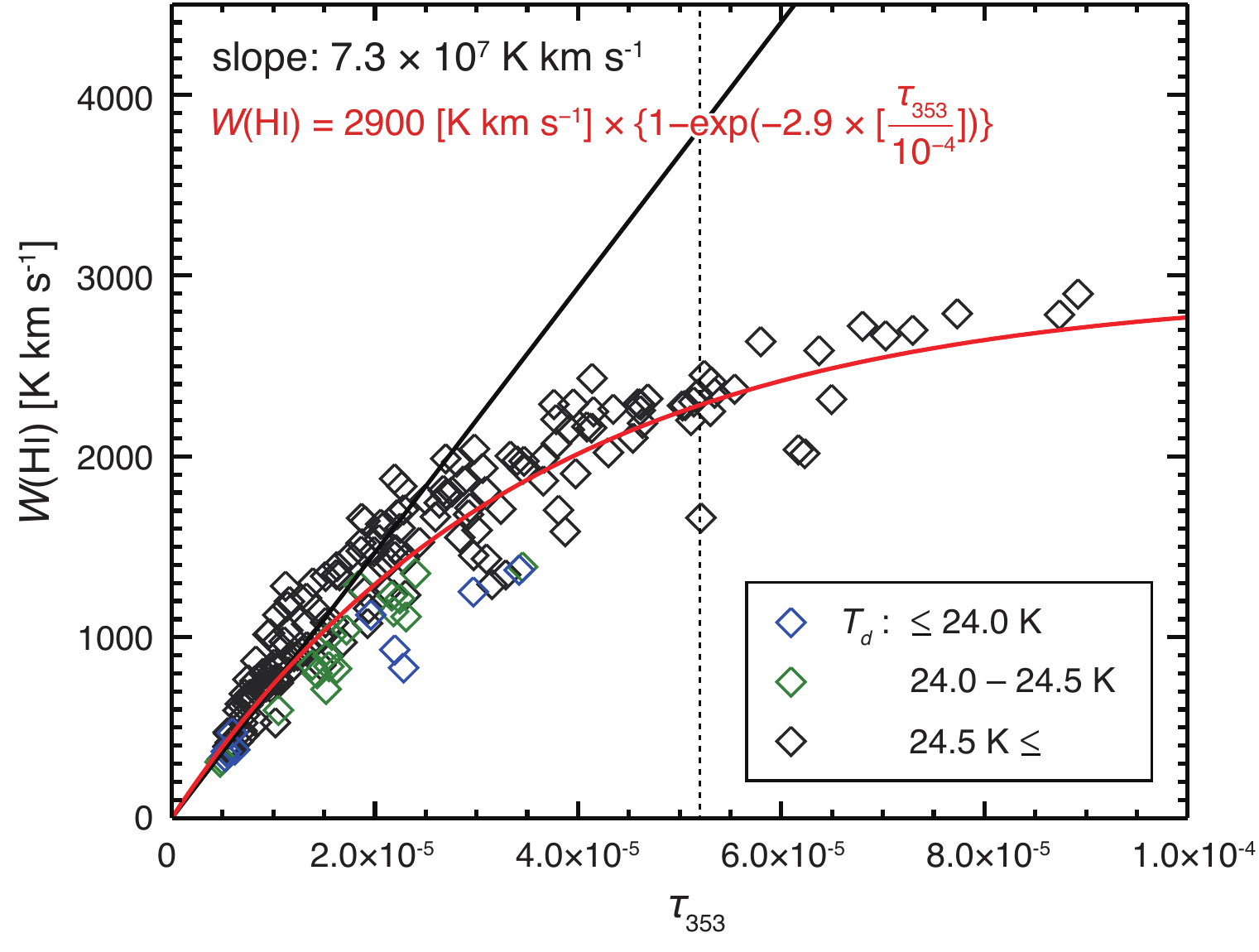}
\caption{Scatter plot between $W$(H{\sc i}) and $\tau$$_{353}$ in the region indicated as white boxes of Figure \ref{fig12}. We show three $T_\mathrm{d}$ range, 24.0 K and lower (blue), 24.0--24.5 K (green), and 24.5 K and higher (black). The black line denotes the result of the least-squares fitting of a straight line whose slope is indicated in the top left corner. The red curve is the function of {Equation (\ref{equ:10}) with $a=2900$ and $b=2.9 \times 10^4$} (see the text). The black dashed line represents $\tau$$_{353}=5.2 \times 10^{-5}$, which is the typical value at the 30 Doradus C region.}
\label{fig13}
\end{center}
\vspace*{2cm}
\end{figure}%

\section*{Appendix}\label{sec:app}
Recent comparative studies between the H{\sc i} and the dust emission confirmed that a large amount of optically thick H{\sc i} gas is associated with high-latitude clouds \citep[e.g., ][]{2014ApJ...796...59F, 2015ApJ...798....6F, 2017ApJ...838..132O,2019ApJ...878..131H,2019ApJ...884..130H}. \cite{2015ApJ...798....6F} concluded that the amount and distribution of the hydrogen gas could be estimated more accurately by using the scaling factor {Equation (\ref{equ:4}) with $X\sim2$}. In this study, we estimate this scaling factor $X$ for the 30~Dor~C region using the integrated intensity of H{\sc i} ($W$(H{\sc i})) and the dust opacity at 353 GHz (${\tau}_{353}$) around 30~Dor~C.

\begin{figure*}
\begin{center}
\includegraphics[width=\linewidth,clip]{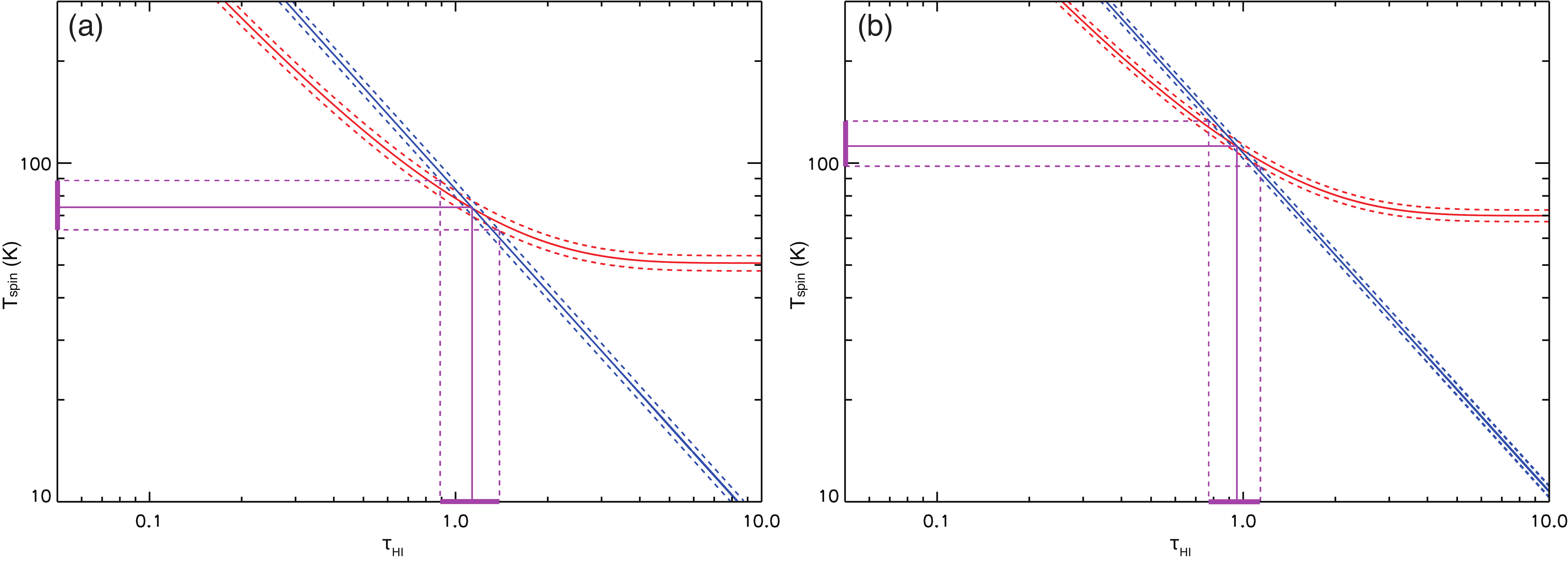}
\caption{Estimations of $T_s$ and ${\tau}_{\mathrm{H}\textsc{i}}$ in the typical cases of (a) $T_s  \sim$ 70 $K$, and  (b) $\sim$110 $K$. Red and blue lines show {E}quations {(\ref{equ:6})} and {(\ref{equ:7})}, respectively. The solution that satisfies both {E}quations {(\ref{equ:6})} and {(\ref{equ:7})} is the crossing point of the two lines. The dashed lines around each line mean the error taking observational parameters into consideration. The purple solid lines represent the solutions of $T_s$ and dashed lines show their errors.}
\label{fig14}
\end{center}
\end{figure*}

\begin{figure*}
\begin{center}
\includegraphics[width=\linewidth,clip]{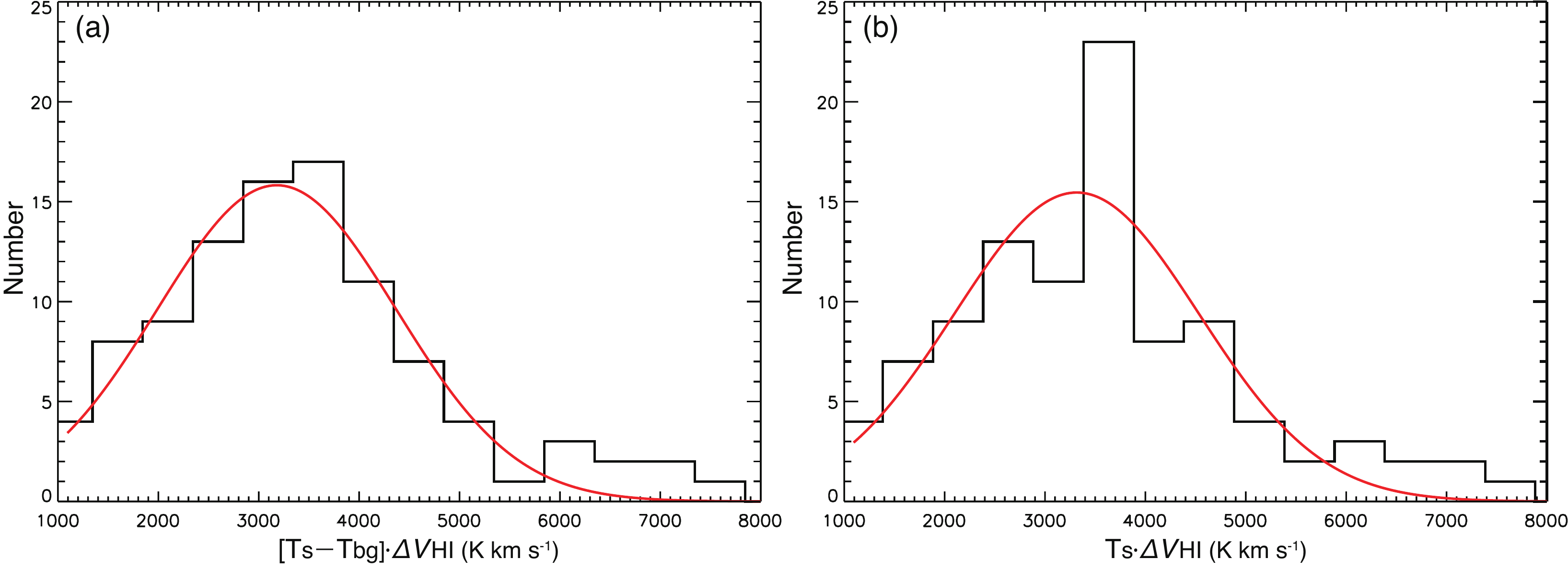}
\caption{{Histograms} of (a) [T$_{\mathrm{s}}-$T$_{\mathrm{bg}}$]$\Delta$$V_{HI}$, and (b) T$_{\mathrm{s}}{\Delta}V_{HI}$ at the region enclosed by white boxes of Figure \ref{fig12}. Red curves indicate the Gaussian functions fitted by these {histograms}. The peak values of these functions are (a) 3200 K km s$^{-1}$, and (b) 3300 K km s$^{-1}$.}
\label{fig15}
\end{center}
\end{figure*}

Figure \ref{fig13} shows a scatter plot between $W$(H{\sc i}) and ${\tau}_{353}$ of the $30\arcmin \times 30\arcmin$ region around 30~Dor~C (see white boxes of Figure \ref{fig12}). $W$(H{\sc i}) is the integrated intensity of H{\sc i}, whose integration velocity range is  from 150 km s$^{-1}$ to 350 km s$^{-1}$. The velocity range contains all interstellar gas of the LMC. We find a non-linear relation when ${\tau}_{353}$ or $W$(H{\sc i}) becomes larger value (see Figure \ref{fig13}). When the dust temperature ($T_d$) is high, spin temperature of H{\sc i} is thought to be also high and hence H{\sc i} gas is expected to be optically thin. We estimated the $N_p$ (H{\sc i}) by the following equation under the assumption that H{\sc i} gas is optically thin,

\begin{equation}
N_p ({\mathrm{H}\textsc{i}})= 1.823 \times 10^{18} \cdot W({\mathrm{H}\textsc{i}}).
\label{equ:5}
\end{equation}
The black line in Figure\ref{fig13} represents the result of the least-squares fits using the data points of $T_d  \geq 24.0$ K and ${\tau}_{353} < 1.5 \times 10^{-5}$ assuming zero intercept and the function is represented as {Equation (\ref{equ:6})},

\begin{equation}
W({\mathrm{H}\textsc{i}})=\tau_{353} \times 7.3 \times 10^7.
\label{equ:6}
\end{equation}
The relation between $N_p$ (H{\sc i}) and ${\tau}_{353}$ is estimated at $T_d  \geq 24.0$ K; 

\begin{equation}
N_p ({\mathrm{H}\textsc{i}})=(1.3 \times 10^{26} ) \times {\tau}_{353},
\label{equ:7}
\end{equation}
where the slope of $1.3 \times 10^{26}$ is calculated by using {E}quations {(\ref{equ:5})} and {(\ref{equ:6})}. As mentioned by \cite{2014ApJ...796...59F}, relation {(\ref{equ:7})} holds as long as the dust properties, such a gas-to-dust ratio, are uniform. In this case, {it is possible to calculate $N_p$(H{\sc i})} from ${\tau}_{353}$ using {E}quation {(\ref{equ:7})}. Then the coupled {E}quations {(\ref{equ:8})} and {(\ref{equ:9})} in the following are used to solve for $T_s$ and ${\tau}_{\mathrm{H}\textsc{i}}$,

\begin{equation}
W({\mathrm{H}\textsc{i}})=[T_s-T_{bg} ] \cdot {\Delta}V_{\mathrm{H}\textsc{i}} \cdot [1-\exp({\tau}_{\mathrm{H}\textsc{i}})],
\label{equ:8}
\end{equation}
\begin{equation}
{\tau}_{\mathrm{H}\textsc{i}}=\frac{N_p({\mathrm{H}\textsc{i}})}{1.823 \times 10^{18}} \cdot \frac{1}{T_s} \cdot \frac{1}{{\Delta}V_{\mathrm{H}\textsc{i}}}.
\label{equ:9}
\end{equation}
Equations {(\ref{equ:8})} and {(\ref{equ:9})} lead to {E}quation {(\ref{equ:10})},

\begin{equation}
W(\mathrm{H}\textsc{i})=a {\cdot} [1-{\exp}(-b {\cdot} {\tau}_{353})],
\label{equ:10}
\end{equation}
where $a=[T_s-T_{bg}] \cdot {\Delta}V_{\mathrm{H}\textsc{i}}$ and $b=\frac{1.3 \times 10^{26}}{1.823 \times 10^{18}} \cdot \frac{1}{T_s} \cdot \frac{1}{{\Delta}V_{\mathrm{H}\textsc{i}}}$. Here, ${\tau}_{\mathrm{H}\textsc{i}}$ is the optically depth, and $T_s$ is the spin temperature. $T_{bg}$ is brightness temperature of background and we assume $T_{bg}=2.7$ K taking the effect of cosmic microwave background into account. ${\Delta}V_{\mathrm{H}\textsc{i}}$ is the H{\sc i} line width given by $W$(H{\sc i})$/$(peak H{\sc i} brightness temperature).

The red curve in Figure \ref{fig13} derived by the least-squares fits using all data points. This function represents the Equation {(\ref{equ:10})} {with} $a=2900$ and $b=2.9 \times 10^4$. Black dashed line in Figure \ref{fig13} indicated the mean value of ${\tau}_{353}=5.2 \times 10^{-5}$ in 30~Dor~C. By adopting the value into curve functions, we derived $W$(H{\sc i}) $=$3800 K km s$^{-1}$ for black line function, and $W$(H{\sc i})$=$2250 K km s$^{-1}$ for red curve function. This result means that the column density of H{\sc i} gas calculated by using {E}quation {(\ref{equ:5})} is estimated to be smaller than that of expected value. Thus, we estimated the scaling factor $X=3800/2250 \sim 1.7$ in the {E}quation {(\ref{equ:4}).}

Using {E}quations {(\ref{equ:8})} and {(\ref{equ:9})}, $T_s$ was solved by the bisection method at each data point. The two relations in the $T_s$--${\tau}_{\mathrm{H}\textsc{i}}$ plane for two typical value of (a) $T_s  \sim 70$ K and (b) $T_s  \sim 110$ K are shown in Figure \ref{fig14}. Figure \ref{fig14} also shows the errors in $T_s$ and ${\tau}_{\mathrm{H}\textsc{i}}$ which were estimated from the 1$\sigma$ noise level of the observation data of H{\sc i} and ${\tau}_{353}$. We estimate the error ranges in Figure \ref{fig14} as follows : (a) $T_s  \sim 70^{+20}_{-13}$ K, ${\tau}_{\mathrm{H}\textsc{i}} \sim 1.14^{+0.25}_{-0.24}$, (b) $T_s  \sim 110^{+25}_{-18}$ K, ${\tau}_{\mathrm{H}\textsc{i}} \sim 0.96^{+0.18}_{-0.18}$. The equations give the solutions with $\sim$20--30 $\%$ errors. Figure \ref{fig15} shows the histogram of (a) $[T_s-T_{bg}] \cdot {\Delta}V_{\mathrm{H}\textsc{i}}$ and (b) $T_s \cdot {\Delta}V_{\mathrm{H}\textsc{i}}$ given by using the $ {\Delta}V_{\mathrm{H}\textsc{i}}$ and the solutions of $T_s$ at each data point. The red curves in Figures \ref{fig15}(a) and (b) represent the results of the Gaussian fitting using those histograms. We estimate $a=3200$ and $b=2.2 \times 10^4$ from these mean values of the Gaussian functions. When $a=3200$ and $b=2.2 \times 10^4$ are substituted into {E}quation {(\ref{equ:10})}, $W$(H{\sc i})$=$2200 K km s$^{-1}$ by adopting ${\tau}_{353}=5.2 \times 10^{-5}$ which is the mean value in 30~Dor~C. In this case, the scaling factor $X \sim 1.7$ in the {E}quation {(\ref{equ:4})} as same as the estimation by red curve function in Figure \ref{fig13}. Accordingly, we estimated the amount of the H{\sc i} gas more accurately using the scaling factor $X \sim 1.7$ in the {E}quation {(\ref{equ:4})} in the present study.

\end{document}